\documentclass[twocolumn,twocolappendix]{aastex63}
\usepackage[utf8]{inputenc}
\usepackage{placeins}
\usepackage{graphicx}
\usepackage{amsmath}
\usepackage{physics}
\usepackage{mathtools}

\usepackage{dcolumn}
\newcolumntype{d}[1]{D{.}{.}{#1}}

\begin{document}

\title{Mean Mass Density near the Sun from the Divergence Theorem and Pulsar Accelerations}

\author[0000-0002-7746-8993]{Thomas Donlon II}
\affiliation{Department of Physics and Astronomy, University of Alabama in Huntsville, 301 North Sparkman Drive, Huntsville, AL 35816, USA}
\correspondingauthor{Thomas Donlon II}
\email{thomas.donlon@uah.edu}

\author[0000-0001-6211-8635]{Lawrence M. Widrow}
\affiliation{Department of Physics, Engineering Physics and Astronomy, Queen’s University, Kingston, ON K7L 3N6, Canada}

\author[0000-0001-6711-8140]{Sukanya Chakrabarti}
\affiliation{Department of Physics and Astronomy, University of Alabama in Huntsville, 301 North Sparkman Drive, Huntsville, AL 35816, USA}

\begin{abstract}
We introduce a new, non-parametric method for estimating the mass enclosed within a sphere of arbitrary radius centered on the Sun. The method is based on the divergence theorem as applied to measurements of the line-of-sight accelerations of millisecond pulsars. We describe a procedure for inferring the mean mass density within a sphere of a given radius centered on the Sun and find results that are consistent with previous analyses. When combined with a model for the distribution of baryons, this provides the mean mass density of dark matter as a function of distance from the Sun, rather than a single value as is typically reported by kinematic studies. However, with the present pulsar data, the method cannot unambiguously measure a signal from the local distribution of dark matter at this time; such a measurement is expected to soon become possible as the amount of pulsar acceleration data grows and its precision improves. We derive an extension of the well-known shell theorem to a spherical-harmonics expansion of the density and potential, and use the result to obtain estimates for density asymmetries with respect to the Galactic midplane from the observed acceleration data.  The predicted asymmetries do not follow the observed distribution of MW disk stars or gas; this can potentially be explained by a non-uniform distribution of dark matter in the Solar neighborhood. \\\vspace{0.5cm}
\end{abstract}

\section{Introduction} \label{sec:intro}

The challenge in determining the distribution of mass in the Milky Way (MW) is that we are generally unable to directly measure the accelerations of gravitational tracers such as stars and gas. Were we able to do so, we could construct a smooth map of the acceleration field and then infer the mass density via Poisson's equation. Instead, we must supplement position and velocity measurements with assumptions about the dynamical state of the Galaxy \citep{BlandHawthornGerhard2016}. For example, with Jeans modeling we make the assumption that the Galactic potential is time-independent and that the bulge and stellar disk, which provide both significant contributions to the mass of the Galaxy and suitable populations of tracers, are in dynamical equilibrium.

Each of these stellar components can be described by a phase space distribution function (DF) $f$, which obeys the collisionless Boltzmann equation \citep[CBE, see][]{BinneyTremaine2008}. For equilibrium systems, $f$ depends on the phase space coordinates only through the integrals of motion (Jeans' theorem). This condition is enough to ensure that we can uniquely determine the potential from measurements of positions and velocities \citep{BinneyTremaine2008, An2021}. Alternatively, we can use velocity moments of the time-independent CBE, namely the continuity and Jeans equations, to determine the potential. Jeans methods have been used for nearly a century to determine the vertical force and the surface density $\Sigma$ as functions of the position $z$ relative to the Galactic midplane \citep{Oort1932, KuijkenGilmore1991, BovyTremaine2012, Garbari2012, Zhang2013, McKee2015, Buch2019}.

In general, the assumption that the Galaxy is in equilibrium is paired with the assumption that it is axisymmetric and symmetric about the midplane. However, the observed density, bulk velocity, and velocity dispersion profiles for disk stars show asymmetries with respect to the Galactic midplane \citep{Widrow2012, YannyGardner2013, Carlin2013, Williams2013,  Xu2015, BennettBovy2019}. The interpretation is that the stellar disk near the Sun is in a state of disequilibrium.  Another manifestation of disequilibrium is the Gaia snail, a spiral-like feature in the number density of stars across the $z-v_z$ plane, which is strongly suggestive of incomplete phase mixing of a perturbation to the disk \citep{Antoja2018}. These departures from the assumption of equilibrium can lead to systematic errors in Jeans modeling \citep{Banik2017,HainesDonghia2019,SivertssonRead2022}. Indeed, disequilibrium is often revealed through Jeans modeling; for example, the residuals of Jeans method models for the DF and potential provide clearly show the Gaia snail and similar features as functions of angle-action variables \citep{LiWidrow2021, LiWidrow2023}. Although there are ways of accounting for some disequilibrium features in Jeans modeling \citep[for example, see Section 5.3.2 of][]{Guo2020}, this often consists of adding additional terms of a series expansion, which is ultimately still an approximation.

In principle, data from surveys such as Gaia \citep{GaiaDR3} and APOGEE \citep{APOGEE} allow for redundancy checks in Jeans modeling for the surface density and vertical force.  Different populations of stars (as defined, for example, by abundance ratios) typically have different vertical structures, which are related to their distribution at birth and subsequent dynamical processes such as vertical heating \citep{RixBovy2013,HagenHelmi2018}. However, so long as the Galaxy is in equilibrium, when taken as tracers of the potential, these distinct populations should give the same results for the vertical force. Recently, \cite{Cheng2024} found that the vertical force inferred from Jeans modeling of two distinct samples, namely thin and thick disk stars, yielded very different results. They considered several explanations for the discrepancy, including the possibility that disequilibrium manifests itself differently in the thin and thick disks.  This explanation is entirely plausible since dynamically distinct populations will respond differently to gravitational perturbations. These issues make it difficult to determine the correct value for the vertical force from tracer stars alone.

Recently \cite{Buckley2023} and \cite{Lim2025} have developed a promising algorithm for deriving the potential from astrometric data based on normalizing flows, an unsupervised machine learning method that is particularly well-suited to dynamical systems. The basic idea is to adjust the potential so as to minimize $\partial \ln{f}/\partial t$. Otherwise, there are no {\it a priori} assumptions about symmetries of the system. In other words, the model does not rigorously impose time-independence -- rather, it finds the potential that minimizes time-dependence of $f$. When \cite{Lim2025} applied the method to Gaia DR3 data, they found departures in the acceleration field from axisymmetry and symmetry about the midplane, which they attributed to disequilibrium processes.

Recent advances in pulsar observations have introduced a promising alternative to Jeans modeling, which makes use of direct measurements of the Galactic acceleration field. There are now direct acceleration measurements for over 50 pulsars, opening up the possibility for determining the mass density from Poisson's equation without any assumptions about the state of the Galaxy \citep{Chakrabarti2021, Moran2023, Donlon2024, Donlon2025}.
One must, however, confront two significant hurdles. First, with only $\sim 50$ acceleration measurements within 3 kpc, the spatial resolution of the acceleration field is only $1-2\,{\rm kpc}$, even if we set aside the considerable measurement uncertainties in the distance to the pulsars and the acceleration itself. Second, we only have access to $a_{\rm los}$, the component of the acceleration relative to the Sun along the line of sight. In \citet{Chakrabarti2021, Donlon2024, Donlon2025} these issues were handled but using a parametric model for the Galactic potential. The best-fit model was found by comparing model predictions for $a_\mathrm{los}$ with measurements toward each of the sources in the data set.  Although we only use pulsar data in this study, our work here can also be applied to other time-domain techniques for direct acceleration measurements that are expected to become viable in the near future, including extreme-precision radial velocity (EPRV) observations \citep{Chakrabarti2020} and eclipse timing \citep{Chakrabarti2022}.

In this paper, we introduce a novel, non-parametric method for determining the mass within concentric spheres centered on the Sun. The method is based on Gauss's law for gravity, which states that the total mass within a volume $V$ is proportional to the integral over the surface of $V$ ($\partial V$) of the acceleration dotted into the normal of the surface. If $V$ is chosen to be a sphere centered on the Sun, then only $a_{\rm los}$ is required. Furthermore, the spherical averaging via the surface integral helps with the sparsity of the data. The method yields $\bar{\rho}(r)$, the mean density as a function of the distance $r$ from the Sun.  Our approach here generalizes the usual reporting of the local dark matter density, $\rho_{\rm DM}$, that are reported in kinematic studies as a single number, to a function of distance.  This function can also be used to estimate the density of dark matter at various distances from the Sun, as well as surface density as a function of position above the midplane, though the second calculation requires that we take a derivative of $\bar{\rho}$, which amplifies the noise in the data.

Gauss's theorem can be extended to an expansion of the density and potential in spherical harmonics, as derived in Section 4. We show that higher order terms in the expansion allow one to extract information about asymmetries in the mass distribution with respect to the midplane of the Galaxy by computing integrals over $\partial V$ of $a_{\rm los}$, weighted by Legendre polynomials. However, Newton's shell theorem no longer applies and these integrals are sensitive to asymmetries in the mass distribution at arbitrarily large distances from the Sun. One can avoid this problem, but at the cost of taking derivatives of the integrals with respect to $r$.

\section{Data}

\begin{figure}[h!]
    \centering
    \includegraphics[width=\linewidth]{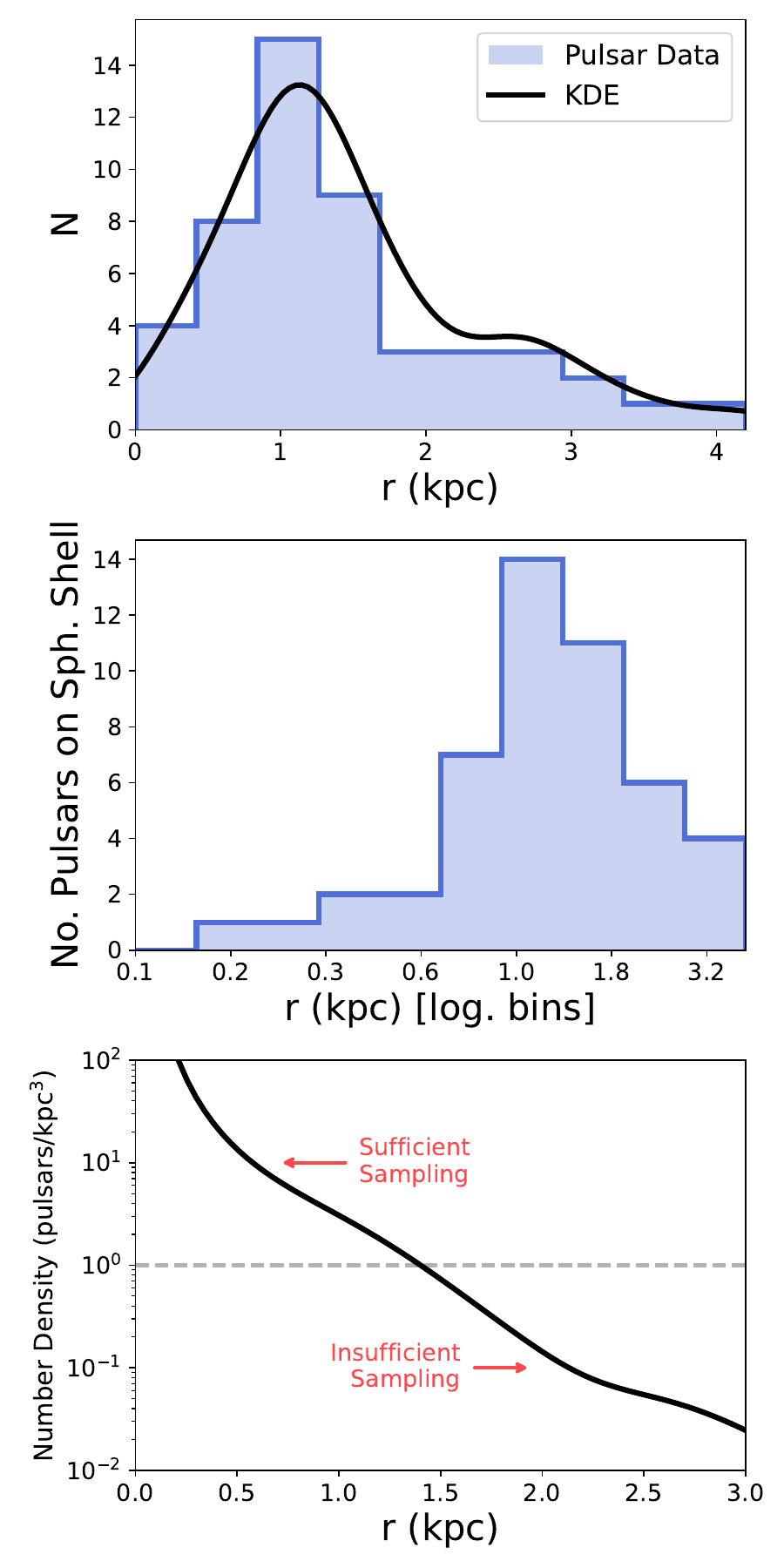}
    \caption{Heliocentric distances of the pulsars in our sample. The black line gives a kernel density estimate of the blue histogram. Inside roughly 1.5 kpc, there is sufficient density of pulsars to obtain a reasonable approximation of the overall acceleration field at a given point. Outside of this distance, the density of the data drops below 1 source per cubic kpc, and the sparseness of the data makes it difficult to effectively interpolate between datapoints.}
    \label{fig:psr_dists}
\end{figure}

The data used here was compiled by \citet[][hereafter D25]{Donlon2025}, and contains line-of-sight accelerations derived from the orbital information of 27 binary millisecond pulsars\footnote{Because the double pulsar J0737-3039A/B shares an acceleration for the two objects, there are only 26 unique binary millisecond pulsar acceleration measurements in the dataset.} plus the spin data of 26 millisecond pulsars. The inferred line-of-sight acceleration for a given pulsar is based on changes to its orbit and spin frequency, which arise from a combination of the Galactic acceleration and other affects. The pulsar data have been corrected for apparent accelerations due to the source's motion on the sky \citep{Shklovskii1970}. For the binary pulsars, we account for the contribution to the orbital decay due to the emission of gravitational waves \citep[e.g.][]{WeisbergHuang2016}. For the pulsars without sufficient binary orbital data, we account for the contribution to the decay of the spin rate due to magnetic braking \citep{Donlon2025}. D25 also removed any sources that have a significant contribution to the acceleration due to an orbital companion. This includes spider pulsars \citep[i.e.][]{CampanaDiSalvo2018}, pulsars which are otherwise interacting with their companion, or pulsars which are being accelerated by a nearby object, such as J2043+1711 \citep{Donlon2024b}. Changes in the observed frequency of a pulsar can possibly be caused by other factors beyond these listed here, although such effects are expected to be small for the vast majority of pulsars \citep[for example, see Table 3 of][]{Donlon2024b}.

In the next section, we show that the contribution from each pulsar in the dataset to the integral in Gauss's law is proportional to $a_\mathrm{los}/r$, where r is the distance of the pulsar to the Sun. We removed outliers from the D25 pulsar sample by eliminating pulsars with $|a_\mathrm{los}|/r > 3\sigma$, where $\sigma$ is the standard deviation of $|a_\mathrm{los}|/r$ across the entire dataset\footnote{This uncertainty includes the intrinsic scatter correction from Appendix B of \cite{Donlon2025}. We emphasize that this is the correct way to treat the uncertainties of the pulsar spindown from that dataset.}. This removes 2 pulsars from our sample -- J0125$-$2327 and J1400$-$1431 -- leaving 50 unique acceleration measurements. The two pulsars with excessively large values of $|a_\mathrm{los}|/r$ overwhelm the algorithm at small radii, where they lead to negative values for the inferred density. It is possible that the large value of $|a_\mathrm{los}|/r$ in J1400$-$1431 is simply due to measurement uncertainties, which are substantial for that pulsar. This is probably not the case for J0125$-$2327, however, which has small measurement uncertainties; it is possible that this pulsar is experiencing a significant peculiar acceleration and/or anomalies in its magnetic field that are leading to its large observed value of $|a_\mathrm{los}|/r$. Identifying the source of this anomaly is outside the scope of this paper, but might be of interest for future studies. 

The results of this work are calculated for both the binary + spin sample, as well as only the 26 binary accelerations. We do this because estimates of accelerations from spin data rely on average magnetic field properties of millisecond pulsars \citep{Donlon2025}. In contrast, the accelerations from binary orbital parameters are directly measured, making the binary data a more precise subset of the data. As a result, it is plausible that the accelerations derived from spin information are less accurate, and could skew the results (as was the case for the 2 removed pulsars). However, we do not observe any major differences between the binary + spin and the binary samples after our quality cut, indicating that this is not a significant problem. 

The heliocentric distance for each pulsar in the D25 catalog is shown in Figure \ref{fig:psr_dists}. As we intend to evaluate the line-of-sight acceleration field at arbitrary points on a sphere of a particular radius, we need a sufficiently dense pulsar sample in order to obtain an accurate interpolation of the true underlying acceleration field. The number density of the pulsar data drops as one looks at distances farther from the Sun; as a result, we can only confidently infer the acceleration field and mean density inside a relatively small volume. The number density drops below 1 source per cubic kpc at a radius of about 1.5 kpc, which provides a rough estimate for the volume inside which our results are meaningful. Similarly, if one bins the pulsars on a logarithmic scale (middle panel of Figure \ref{fig:psr_dists}), the number of pulsars in each bin corresponds to the number of pulsars on the surface of a sphere with radius $r$. The number of pulsars on each shell is small within $r\sim 600$ pc from the Sun, which provides a lower distance bound outside which our results are significant. To summarize, we expect our results to be the most reliable between 0.6 and 1.5 kpc from the Sun due to the spatial distribution of pulsars.

Note that pulsar distances are not easy to measure. The median distance uncertainty across our pulsar catalog is 250 pc with the uncertainties for a several pulsars reaching a few kpc. Distance uncertainties enter or analysis in two ways. First, the connection between pulsar spin-down rates and the Galactic acceleration must be corrected for a kinematic effect due to the tangential motion of the pulsar (the Shklovskii effect, see \citealt{Shklovskii1970}), which depends on the distance. Second, interpretation of the acceleration of a pulsar in terms of models for the Galactic potential requires that we know the position of the pulsar. In the following section we introduce a procedure which weights the contribution of each pulsar based on the uncertainty in its observed acceleration; consequently, the impact of pulsars with very large distance uncertainties on our final results is minimized.

\section{The Divergence Theorem} \label{sec:gauss_law}

\subsection{Mean Volume Density} \label{sec:mean_vol_dens}

Let $\mathbf{a}=\mathbf{a}(\mathbf{x})$ be the gravitational acceleration at position $\mathbf{x}$, $V$ be an arbitrary volume, and $\partial V$ be the surface of $V$ with unit normal $\mathbf{\hat{n}}$. The divergence theorem as applied to Newtonian gravity relates the mass in $V$ to the integral over $\partial V$ of the normal component of $\mathbf{a}$:
\begin{equation} \label{eq:gauss_law}
    \oint_{\partial V} \mathbf{a}(\mathbf{x})\cdot \mathbf{n}\dd A = -4\pi G \int_V \rho(\mathbf{x}) \; \dd V.
\end{equation} In this work, we use heliocentric spherical coordinates $(r,\theta,\phi)$,
where $r$ is the distance from the Sun, $\theta$ is the angle from the north galactic pole, and $\phi$ is Galactic azimuth with $\phi=0$ being the direction towards the Galactic center. If we choose $V$ to be a sphere of radius $r$ centered on the Sun, then the mean density within $V$ is given by
\begin{equation}\label{eq:mean_rho}
\bar{\rho}(r) = -\frac{3}{4\pi G r}\int a_r(r,\theta,\phi) d\Omega
\end{equation}
where $d\Omega = \sin{\theta} d\theta d\phi$, and $a_r(r,\theta,\phi)$ is the line-of-sight acceleration at a given point. 

Pulsar timing provides a direct measurement of the relative acceleration of a pulsar with respect to the Sun -- that is, $a_h(\mathbf{x}),$ where  \begin{align}
    a_h(\mathbf{x}) &= \left(\mathbf{a}(\mathbf{x}) - \mathbf{a}_\odot\right)\cdot \mathbf{\hat{r}} \\ \nonumber
    &= a_r(\mathbf{x}) - \mathbf{a}_\odot \cdot \mathbf{\hat{r}}.
\end{align} Since the integral of ${\bf a}_\odot\cdot \hat{\bf n}$ over the sphere vanishes, we can substitute $a_h$ for $a_r$ in Equation \ref{eq:mean_rho}.



\subsection{Surface Density}


Suppose that the mass density is only a function of $z$ within $V$. The mean density as a function of distance from the Sun is then \begin{equation}
    \bar{\rho}(r) = \frac{3}{4\pi r^3} \int_0^r 2\pi (r^2 - z^2)\rho(z) \dd z,
\end{equation} and therefore \begin{equation} \label{eq:drho_dr}
    \dv{\bar{\rho}(r)}{r} = \frac{3}{r^2} \int_0^r \rho(z) \dd z - \frac{3\bar{\rho}(r)}{r}.
\end{equation} The surface density within a distance $z$ of the midplane is defined as \begin{equation} \label{eq:surf_dens}
    \Sigma(r) = 2\int_0^r \rho(z) \dd z.
\end{equation} We can then use Equation \ref{eq:drho_dr} to obtain an expression for the surface density that depends only on the value and derivative of the mean density as a function of distance from the Sun: \begin{equation} \label{eq:sigma}
    \Sigma(r) = 2 r \bar{\rho}(r) + \frac{2r^2}{3}\dv{\bar{\rho}(r)}{r}.
\end{equation} This expression is only exact in the case that the density profile is plane-symmetric; otherwise, it provides an approximation for $\Sigma$. The general case is discussed in Appendix \ref{app:general}. 

\subsection{Interpolation} \label{sec:interp}

\begin{figure}
    \centering
    \includegraphics[width=\linewidth]{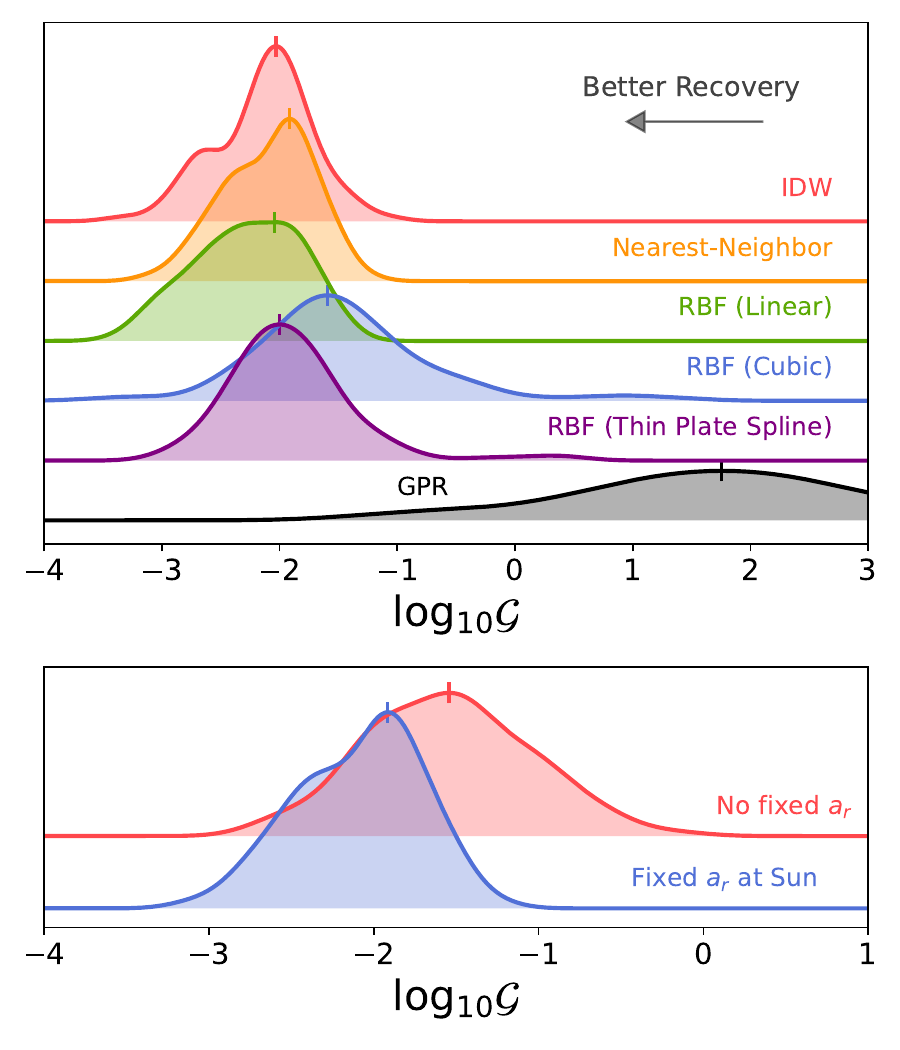}
    \caption{Performance of various interpolation algorithms in recovering the true $\bar{\rho}(z)$ profile from randomly shuffled data. Smaller values of $\log_{10} \mathcal{G}$ correspond to better recovery of the actual underlying accelerations. The distribution of each algorithm is a kernel density estimate over 100 Monte Carlo samples of the observed pulsar data. \textit{Top:} Nearest neighbor interpolation performs the best out of these algorithms, so it is used for the remainder of this paper. \textit{Bottom:} The inclusion of an extra point at the location of the Sun with $a_r = 0$ significantly improves the ability of the interpolation procedure to recover the underlying mean density profile. }
    \label{fig:interp}
\end{figure}

\begin{figure*}
    \centering
    \includegraphics[width=0.6\linewidth]{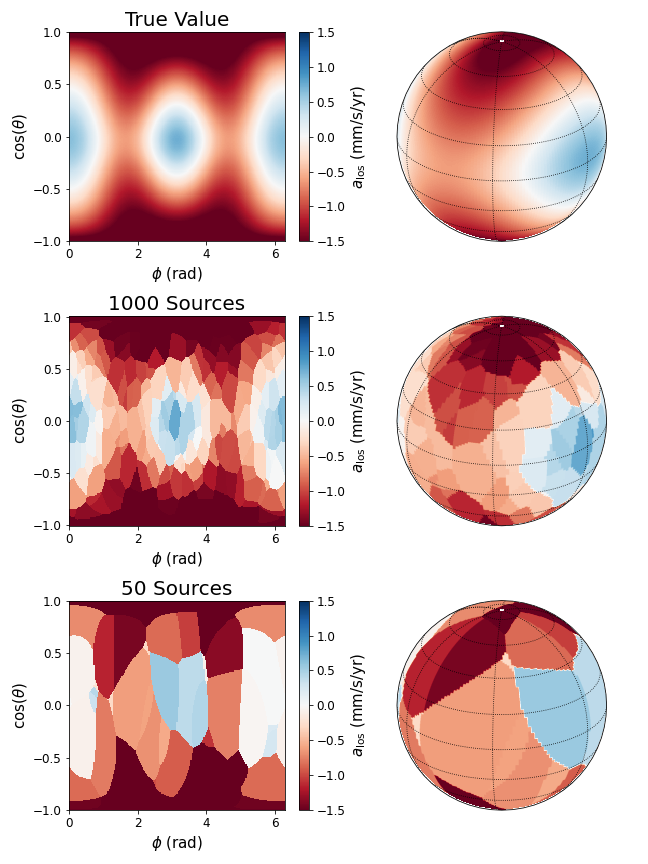}
    \caption{Examples of the line-of-sight acceleration field used in our calculation of the mean volume density. The accelerations shown here are for sphere with radius $r=0.75$ kpc centered on the Sun, calculated using the \textsc{Gala} \textit{MilkyWayPotential2022} model. We show the interpolated on-sky approximation of the acceleration field for different numbers of sources distributed randomly within a cube with side length 2 kpc centered on the Sun. Although the integrand only vaguely resembles the true acceleration field for 50 sources, we find that the integral is still accurate to within 10\% on average. With 1000 sources, the observed acceleration distribution is similar to ground truth, and the integral is accurate to within 1\%.}
    \label{fig:integral}
\end{figure*}


The integral in Equation \ref{eq:mean_rho} requires that we have a model for $a_h$ at all points on a sphere of radius $r$. The choice of interpolation scheme is therefore a key part of the analysis because we have only 50 data points. 

We used mock data to test several interpolation schemes, including inverse distance weighting (IDW), nearest neighbor interpolation (also known as Voronoi tiling), and radial basis function interpolation through Gaussian Process Regression. To construct the mock data, we replaced the measured acceleration at each observation point in the real data set with the acceleration as calculated in \textit{MilkyWayPotential2022}, the standard MW model in the \textsc{Gala} Python package \citep{Gala}. These accelerations were used to calculate $\bar{\rho}_{\rm test}(r)$ via Equation \ref{eq:mean_rho} using the \textsc{SciPy} implementation of each interpolation schemes. We then calculated the dimensionless goodness-of-fit parameter \begin{equation}
    \mathcal{G} = \frac{1}{(r_+ - r_- )} \int_{r_-} ^{r_+} \dd r \left(\frac{\bar{\rho}_\mathrm{gala}(r) - \bar{\rho}_\mathrm{test}(r)}{\bar{\rho}_\mathrm{gala}(r)}\right)^2,
\end{equation} where $\rho_{\rm gala}$ is the true density from {\sc Gala}. The integral extends from $r_-$ = 0.1 kpc to $r_+$ = 2 kpc, which corresponds to the regions where most of the pulsars are located. To improve the statistics, we shuffle the longitudes, latitudes, distances, and uncertainties of each measurement, then repeat the exercise 100 times and compare the distribution of random pulls.


The results of these tests are shown in Figure \ref{fig:interp}. The nearest neighbor interpolation algorithm performed the best, although IDW, linear kernel RBF, and thin plate spline kernel RBF also performed nearly as well as nearest neighbor. We use nearest neighbor interpolation for the remainder of this paper, particularly because it is especially simple and fast compared to the other algorithms. 

In applying the nearest neighbor algorithm to the observed pulsar data, we replaced the distance between the test point and the observation point of the $i$'th pulsar, $d_i$, with an effective distance 
\begin{equation}
    d_{\mathrm{eff},i} = d_i/\sigma_{a_\mathrm{los},i}^{1/p},
\end{equation} that is meant to account for measurement uncertainties in $a_\mathrm{los}$. Here $p$ is a tunable parameter that sets how strongly one weights the relative uncertainties of the acceleration data. For $p\gg 1$, we are essentially ignoring measurement uncertainties whereas for $p\ll 1$, the entire volume is assigned the line-of-sight acceleration of the data point with the smallest $\sigma_{a_\mathrm{los}}$. In what follows we set $p=5$ though our results are insensitive to $p$ in the range $3-10$.

An example of the on-sky integrand for $\bar{\rho}(r)$ at $r=750$ pc is shown in Figure \ref{fig:integral}. We also include an example of what the integrand might look like with 1000 measurements. The integral was computed numerically using uniform spacing in $\phi$ and $\cos\theta$, where the grid had 600 steps in $\phi$ and 300 steps in $\cos\theta$ (for a step size of roughly 1/100 radians). Remarkably, even though the integrand with 50 sources only vaguely resembles the true integrand, the integral is fairly accurate with $\bar{\rho}$(750 pc) determined to within about 10\%. With 1000 sources, the accuracy improves to better than 1\%.

\subsection{Behavior Near the Sun}

As shown in Figure \ref{fig:psr_dists}, there are not very many pulsars in our dataset located within a few hundred pc of the Sun. Due to this lack of data, our method struggles to properly recover the accelerations at small $r$. Whichever pulsar is closest to the Sun ends up dominating the interpolated $a_r$ field near the Sun, which causes the estimated value of $\bar{\rho}(r)$ to be incorrect for distances smaller than a few hundred pc. However, as discussed in Section \ref{sec:mean_vol_dens}, we have the mathematical constraint that $a_r$ must vanish at $r=0$. We therefore include an artificial data point with $a_r$ = 0 at the location of the Sun throughout the rest of this paper. This procedure leads to a dramatic reduction in $\log \mathcal{G}$, as seen in the bottom panel of Figure \ref{fig:interp}.

\subsection{Results}

\begin{figure}
    \centering
    \includegraphics[width=\linewidth]{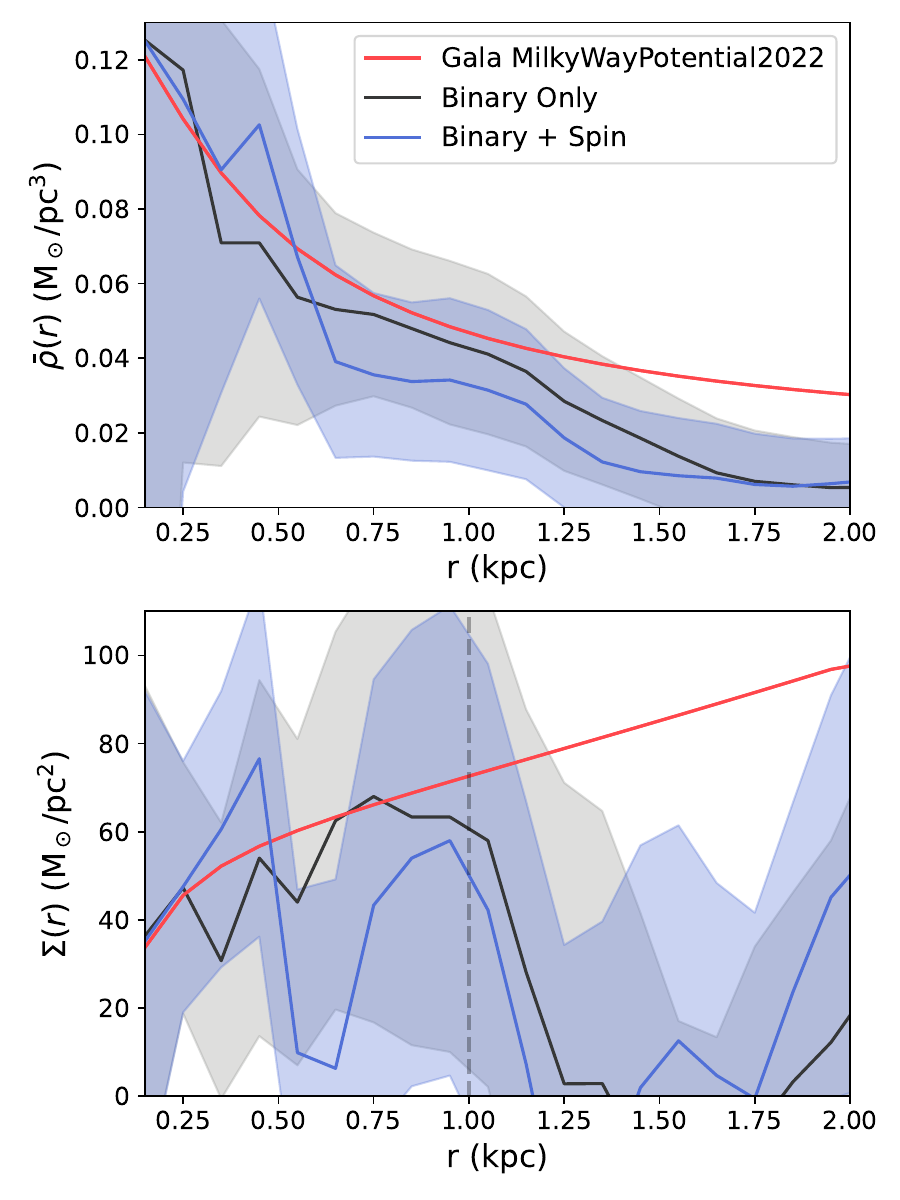}
    \caption{Mean volume and surface density as a function of distance from the Sun, inferred from the pulsar data (blue, with $1\sigma$ uncertainty shaded region). A reference density model fit to observed MW data from the \textsc{gala} Python package is shown in red. The inferred volume and surface densities agree with the kinematic model until about 1.25 kpc from the Sun, when both densities become larger than the simple kinematic predictions.  }
    \label{fig:data_vs_model}
\end{figure}

In the top panel of Figure \ref{fig:data_vs_model}, we show the spherically-averaged density as a function of distance from the Sun. A 1$\sigma$ uncertainty region is also provided, which was calculated as the standard deviation of $\bar{\rho}(r)$ over 100 random realizations of the pulsar data. We assume here that the uncertainties in $a_h$ and parallax (rather than distance) are normally distributed. The overall trends are similar to those expected for models of the Galactic mass distribution that are fit to kinematic observations. In general, the density decreases with $r$, and the curve flattens at large $r$. Outside of 1.1 kpc where the sources become sparse, the observed mean density rapidly drops towards zero, indicating that we are probably not recovering the true behavior at these distances.

In the bottom panel, we show $\Sigma$ as a function of distance from the mid plane. Within 1.1 kpc, the results are broadly consistent with the models of the MW surface density from kinematic data, although the uncertainty in $\Sigma$ is larger than that for $\bar{\rho}$ due to numerical noise generated from the derivative in Equation \ref{eq:sigma}. There is a dip in the inferred surface density at about 0.6 kpc from the Sun in the binary + spin data, although the significance of this dip is only about 1$\sigma$ and does not appear when only the binary data is used. Beyond 1.1 kpc from the Sun, the surface density takes on unphysical values; once again, this is related to the sparsity of data points at those distances. 



It is important to point out that this is a model-independent result; we have not assumed any particular form for the potential or density of the MW in order to carry out this analysis. This is a substantial advantage for the pulsar data over kinematic methods such as Jeans analysis, which either have to assume that the Galaxy is in dynamical equilibrium and/or select a particular analytical form for Galactic features like the density of stars, the vertical force profile, or the velocity field of the Galaxy. This is not to say that our method doesn't make any assumptions -- we are assuming that (i) the acceleration of each pulsar is only due to the Galactic potential, and that (ii) any density features in the Galaxy are large enough that we see them in our relatively sparse spatial sampling. However, both of these assumptions will become more valid over time as more acceleration data becomes available, and any oddities in the accelerations of individual pulsars are discovered and eliminated. 

\subsection{Power Law Fit}

\begin{figure}
    \centering
    \includegraphics[width=\linewidth]{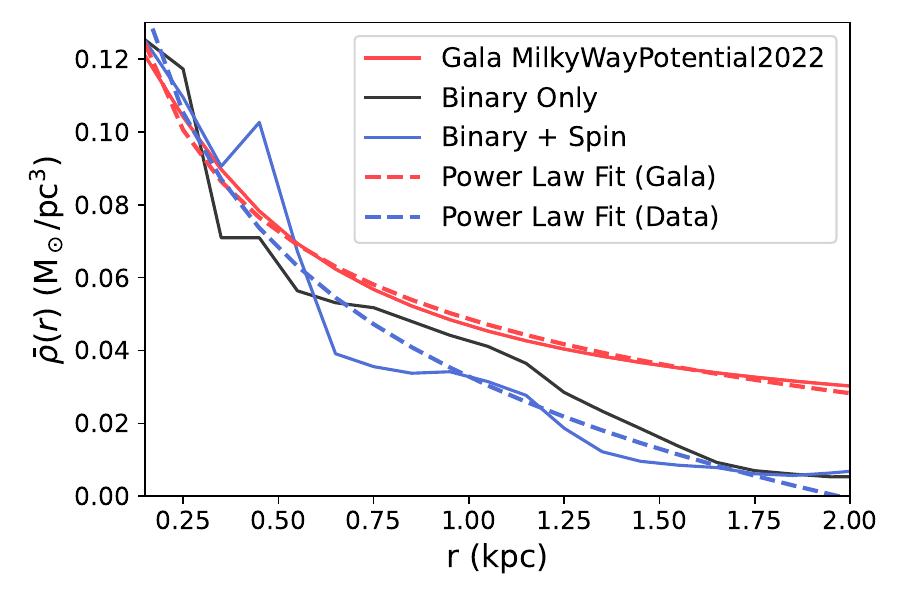}
    \caption{Power law fits to the \textsc{Gala} \textit{MWPotential2022} model and the observed pulsar data. The pulsar data is fit by a steeper slope, potentially implying a disk that is more vertically compact than in the kinematic model.}
    \label{fig:power_law}
\end{figure}

Although our method is non-parametric in principle, fitting a parametric model to the results may aid in comparisons with other methods or datasets. The mean density as calculated from the \textsc{Gala} potential and the observed pulsar data  approximately follows a power law as a function of distance from the Sun. For convenience, we have fit a power law of the form \begin{equation} \label{eq:power_law}
    \bar{\rho}(r) = (\rho_\odot - \bar{\rho}_{+}) [r/\mathrm{kpc}]^{-\alpha} + \bar{\rho}_{+}
\end{equation} to both cases. Physically, the parameter $\bar{\rho}_{+}$ corresponds to the mean density within a spherical volume with a large radius centered on the Sun, $\rho_\odot$ corresponds to the density at the position of the Sun, and the parameter $\alpha$ sets the steepness of the power law. As the Sun is located close to the disk, $\rho_\odot\approx\rho_0$, the density in the midplane. The \textsc{Gala} model has fit parameters $\rho_\odot=0.048$ M$_\odot$/pc$^3$, $\alpha=0.23$, and $\bar{\rho}_+=-0.089$ M$_\odot$/pc$^3$. The fit to the data has optimal parameters $\rho_\odot=0.026\pm0.002$ M$_\odot$/pc$^3$, $\alpha=0.77\pm0.23$, and $\bar{\rho}_+=-0.027\pm0.016$ M$_\odot$/pc$^3$. It is clear that the pulsar data implies a steeper slope than the \textsc{Gala} kinematic model, which would correspond to a disk that is less vertically extended than suggested by the kinematic models. The negative value of $\bar{\rho}_{+}$ in both fits may be an indication that a single power law is not a sufficient model for the mean density distribution.

\subsection{Radial Determination of Dark Matter Density}

\begin{figure}
    \centering
    \includegraphics[width=\linewidth]{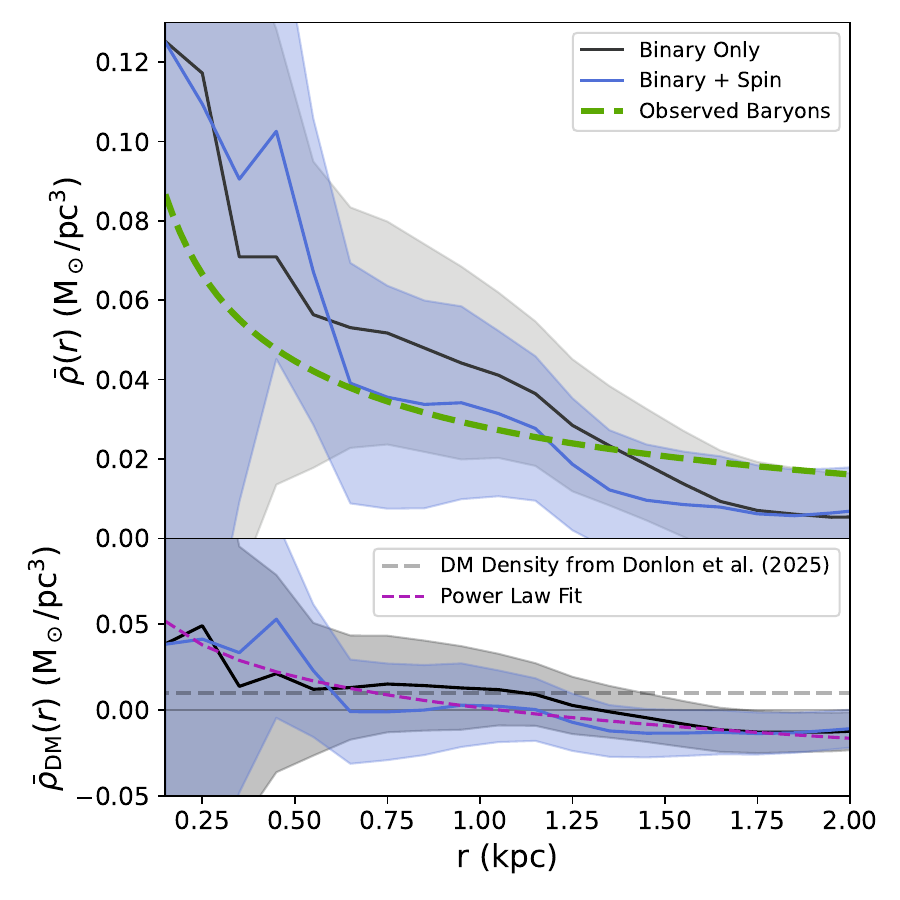}
    \caption{Estimate of the mean density of dark matter within a sphere of radius $r$. The top panel shows the total mean density inferred by the pulsar data compared to the a model for the mean density of baryons in the local solar neighborhood. The bottom panel shows the residual of the mean density inferred from the pulsar data minus the observed baryon distribution, which corresponds to an estimate of the mean dark matter density. The mean dark matter density appears to decrease as a function of $r$, although the data are also consistent with a uniform dark matter density out to 1 kpc. This method infers a negative dark matter density beyond 1 kpc from the Sun, which is a sign that our procedure is underestimating the mean density at those distances (probably due to the low density of pulsars at large $r$). }
    \label{fig:dm_dens_fr}
\end{figure}

It is possible to use a measurement of $\bar{\rho}(r)$ to determine the mean dark matter density enclosed within a sphere of radius $r$. A determination of the mean dark matter density as a function of radius is valuable, because kinematic studies are often restricted to measuring the surface density at a specific height from the midplane rather than a range of heights. As a result, these kinematic studies are forced to assume that the density of dark matter is uniform within the probed volume, although there is no physical reason why this must be the case. 

Given a model for the baryonic matter near the Sun, $\bar{\rho}_\mathrm{bary}(r)$, then the corresponding dark matter density can be calculated as \begin{equation}
    \bar{\rho}_\mathrm{DM}(r) = \bar{\rho}_\mathrm{tot}(r) - \bar{\rho}_\mathrm{bary}(r).
\end{equation} This requires a model for $\bar{\rho}_\mathrm{bary}(r)$, which can be obtained from observations of stars and gas near the Sun. As a demonstration, we model the distribution of baryons in the MW as three $\sech^2$ disks; a gas disk with a scale height of 50 pc, a thin disk with a scale height of 280 pc, and a thick disk with a scale height of 800 pc. These disks are normalized to the relative stellar disk amplitudes of \cite{BennettBovy2019}, along with the surface densities of each component determined by \cite{McKee2015}.

The model for $\bar{\rho}_\mathrm{bary}(r)$ is plotted alongside the observed pulsar data in Figure \ref{fig:dm_dens_fr}. The residual of the pulsar data and the baryon model then provides an estimate of $\bar{\rho}_\mathrm{DM}(r)$. For comparison, we also plot the local density of dark matter as determined by fitting potential models to pulsar accelerations by \cite{Donlon2025} as a dashed gray line. The pulsar data appears to suggest that the density of dark matter is higher close to the Sun than farther away, which could be indicative of a thin disk of dark matter within roughly 0.6 kpc of the midplane (corresponding to an exponential disk with scale height $\lesssim$0.3 kpc, similar to that of the thin disk), or dark substructures as we discuss below. 

Fitting the power law model from Equation \ref{eq:power_law} to the density of dark matter produces the best-fit parameters $\rho_\odot = 0.000\pm0.003$ $M_\odot$/pc$^3$, $\alpha$ = 0.46 $\pm$ 0.37, and $\bar{\rho}_+$ = $-0.04\pm0.04$ $M_\odot$/pc$^3$. This value of $\alpha$ is tentative evidence for a nonuniform distribution of dark matter near the Sun. However, if we instead only fit the power law to the data within 1 kpc where the density of pulsars is high, the best-fit parameters become $\rho_\odot = 0.006\pm0.012$ $M_\odot$/pc$^3$, $\alpha$ = 0.4 $\pm$ 1.5, and $\bar{\rho}_+$ = $-0.04\pm0.27$ $M_\odot$/pc$^3$. This result is generally consistent with a uniform distribution of dark matter within 1 kpc (in other words, $\alpha=0$), and we cannot yet differentiate between a uniform or nonuniform distribution of dark matter given the current uncertainties in the pulsar data. 

\section{Multipole Expansion} \label{sec:multipole}

In the previous section, we used the mathematical statement that the average of $a_r$ on a sphere centered on the Sun is proportional to the average density inside the sphere to infer $\bar{\rho}$ as a function of heliocentric radius. In this section, we show that the analysis can be extended to averages of $a_r$ and $\rho$ weighted by spherical harmonics. 

We begin by writing the gravitational potential near the Sun as a spherical harmonics expansion; \begin{equation}
    \Phi(r,\Omega) = \sum_{l,m} \Phi_{lm}(r) Y_{lm}(\Omega),
\end{equation} where $Y_{lm}(\Omega)$ are the usual spherical harmonics and\begin{equation}
    \Phi_{lm}(r) = \int \dd \Omega \; \Phi(r,\Omega) Y^*_{lm}(\Omega).
\end{equation} Similarly, we have spherical harmonics expansions for $\rho$ and $a_r$. Poisson's equation yields the following relation between $\Phi_{lm}$ and $\rho_{lm}$: \begin{align}
    \Phi_{lm}(r) = -\frac{4\pi G}{2l + 1} \Bigg( \frac{1}{r^{l+1}} \int_0^r \dd s\; s^{l+2} \rho_{lm}(s) \\ \nonumber
    +\; r^l \int_r^\infty \frac{\dd s}{s^{l-1}}\rho_{lm}(s) \Bigg),
\end{align} (see, for example, \citealt{BinneyTremaine2008}). The spherical harmonic coefficients for the radial acceleration are then \begin{align} \label{eq:a_r_def}
    a_{r,lm}(r) = -\frac{4\pi G}{2l + 1} \Bigg( \frac{l+1}{r^{l+2}} \int_0^r \dd s\; s^{l+2} \rho_{lm}(s) \\ 
    - \;l r^{l-1} \int_r^\infty \frac{\dd s}{s^{l-1}}\rho_{lm}(s) \Bigg). \nonumber 
\end{align}

When $l=m=0$, we recover the divergence theorem. We begin by noting that \begin{align}
    a_{r,00}(r) &= \int \dd \Omega\; a_r(r,\Omega) Y^*_{00}(\Omega) \\ \nonumber 
                &= \frac{1}{\sqrt{4\pi}}\int \dd \Omega\; a_r(r,\Omega) \\ \nonumber 
                &= \sqrt{4\pi} \; \langle a_r \rangle_\mathrm{sph},
\end{align} where $\langle \dots \rangle_\mathrm{sph}$ indicates the average of a quantity across the surface of a sphere centered on the Sun. On the other hand, from Equation \ref{eq:a_r_def}, we have \begin{align} \label{eq:tmp1}
    a_{r,00}(r) &= - \frac{4\pi G}{r^{2}} \int_0^r \dd s\; s^{2} \rho_{00}(s) \\ \nonumber
                &= - \frac{\sqrt{4\pi} G}{r^{2}} \int_0^r \dd s\; s^{2} \int \dd \Omega\; \rho(s,\Omega),
\end{align} and because the double integral in the final line of Equation \ref{eq:tmp1} corresponds to the enclosed mass within $r$, we recover \begin{equation}
    \langle a_r \rangle_\mathrm{sph} = -\frac{G M_\mathrm{enc}(r)}{r^{2}}.
\end{equation}

\subsection{Measures of Density Asymmetry}

\begin{figure}
    \centering
    \includegraphics[width=\linewidth]{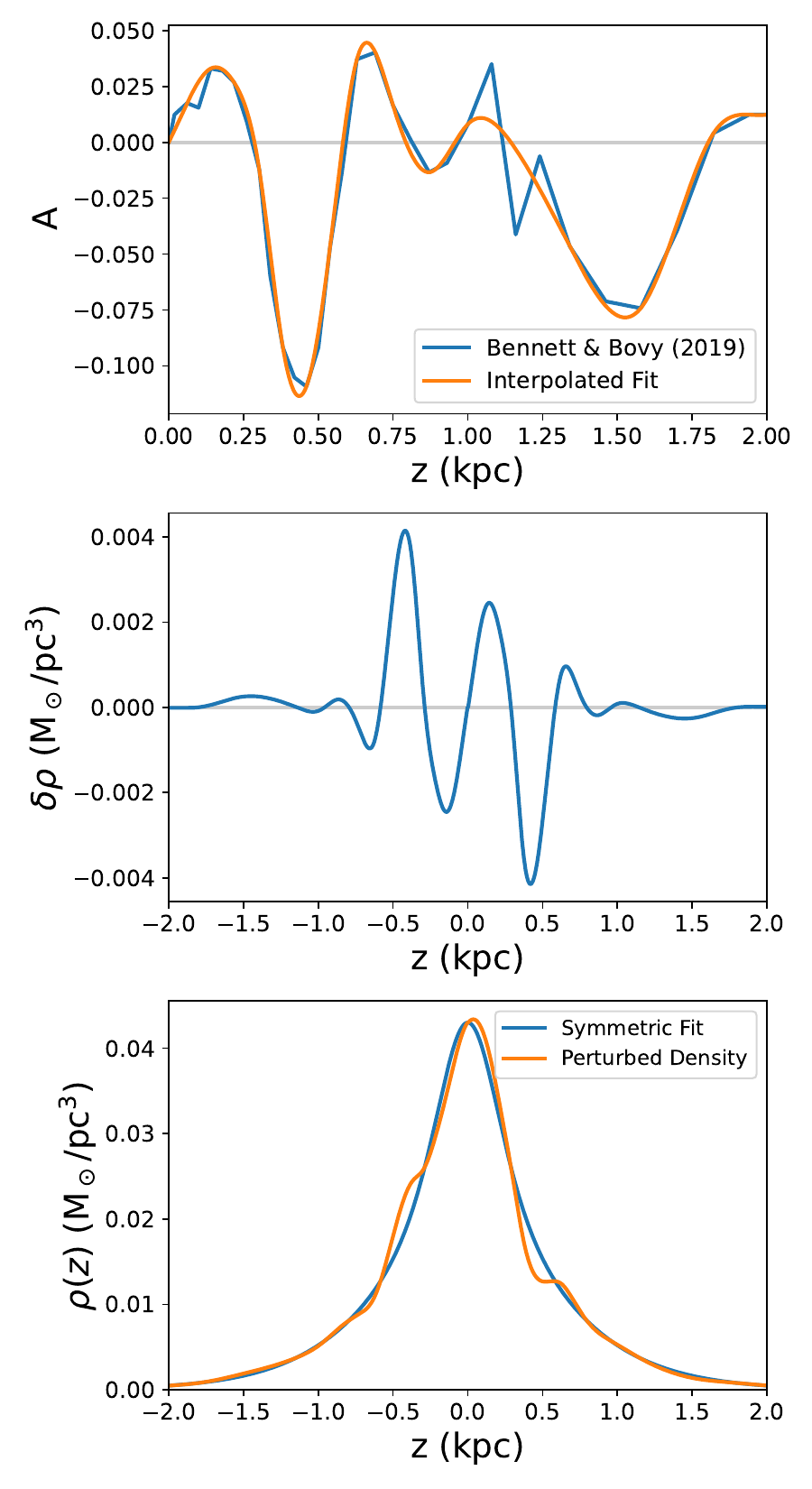}
    \caption{The North-South disk density asymmetry. \textit{Top:} The asymmetry parameter $A$, as measured by \cite{BennettBovy2019}. An interpolated fit is plotted on top of the observed values, which is used to obtain a smooth density in the lower plots. \textit{Middle:} The density asymmetry $\delta\rho$ as a function of vertical height. \textit{Bottom:} The perturbed (actual) density of disk stars, plotted on top of a symmetric douple-exponential disk to emphasize the size of the density perturbations. }
    \label{fig:disk_asym}
\end{figure}

This framework produces further results when one considers terms of the spherical harmonic expansion that are not spherically symmetric. We consider the case $l=1$, $m=0$ where \begin{align} \label{eq:a_r_10_def}
    a_{r,10}(r) &= \int \dd \Omega \; a_r(r,\Omega) Y^*_{10} \\ \nonumber
             &= \sqrt{12\pi} \; \langle a_r \cos \theta \rangle_\mathrm{sph}(r),
\end{align} which is a measure of an asymmetry in the acceleration with respect to the midplane. We note that \begin{align}
    \langle a_r \cos \theta \rangle_\mathrm{sph}(r) &= \langle a_h \cos \theta \rangle_\mathrm{sph}(r) - a_{\odot,z}\langle \cos^2 \theta \rangle_\mathrm{sph}(r) \\ \nonumber 
    &= \langle a_h \cos \theta \rangle_\mathrm{sph}(r) - \frac{a_{\odot,z}}{3},
    \end{align} where $a_h$ is again the directly measured quantity from the pulsars (which includes the Solar acceleration), and $a_{\odot,z}$ is the vertical component of the Sun's acceleration. 

If $a_r$ is identical above and below the midplane at radius $r'$, then $\langle a_r \cos \theta \rangle_\mathrm{sph}(r') = 0$. Any non-zero values of this quantity therefore correspond to differences in the acceleration profile above and below the disk. 

The quantity $a_{r,10}(r)$ can be related to the local density through Equation \ref{eq:a_r_def}, \begin{align}
    a_{r,10}(r) = -\frac{4\pi G}{3} \Bigg( \frac{2}{r^{3}} \int_0^r \dd s\; s^{3} \rho_{10}(s) \\ \nonumber 
    -\; \int_r^\infty \dd s\; \rho_{10}(s) \Bigg).
\end{align} We see that for $l\geq1$, the $Y_{lm}$-weighted average of $a_r$ over the sphere includes a contribution from outside the sphere, which makes it difficult to physically interpret. However, its derivative with respect to $r$ depends only on the density interior to the sphere: \begin{equation} \label{eq:tmp3}
    \dv{\langle a_h \cos\theta \rangle_\mathrm{sph}}{r} = - \sqrt{\frac{4\pi}{3}}G \left( 
     \rho_{10}(r) 
    - \frac{2}{r^{4}} \int_0^r \dd s\; s^{3} \rho_{10}(s)
    \right),
\end{equation} where, explicitly: \begin{equation} \label{eq:shell_ext}
    \rho_{10}(r) = \sqrt{\frac{3}{4\pi}}\int \dd \Omega \; \rho(r,\Omega) \cos\theta.
\end{equation} We point out that Equation \ref{eq:tmp3} is written in terms of $a_h$ rather than $a_r$, as this quantity no longer depends on the Solar acceleration. This is a generalization of the well-known shell theorem for spherically-symmetric mass distributions.

This quantity can also be expressed a function of density rather than $\rho_{10}$. To begin, we compute the first term on the right-hand side of Equation \ref{eq:tmp3}: \begin{align} \label{eq:tmp4}
    \sqrt{\frac{4\pi}{3}}\rho_{10}(r) &= \int_0^{2\pi} \dd \phi \int_{-1}^1 \dd \cos\theta \; \cos\theta \rho(r, \theta,\phi) \\ \nonumber
     &\approx \frac{2\pi}{r^2} \int_{0}^r \dd z \; z \rho(z),
\end{align} where in the second line, we replace $\rho(r,\theta,\phi)$ by a plane-symmetric approximation to $\rho$. 

Following a similar procedure, we can write the second term on the right-hand side as \begin{align}
    & \sqrt{\frac{4\pi}{3}}\frac{2}{r^4}\int_0^r \dd s \; s^3\rho_{10}(s) = \\ \nonumber
    &= \frac{2}{r^4} \int_{-r}^r \dd z \; z \int_{0}^r \dd R\; R \int_0^{2\pi} \dd \phi \; \rho(R,\phi,z) \\ \nonumber
    &\approx \frac{2\pi}{r^4} \int_{0}^r \dd z\; z (r^2 - z^2) \rho(z).
\end{align} The first term of this expression cancels out Equation \ref{eq:tmp4}, leaving \begin{align} \label{eq:the_great_relation}
    \dv{\langle a_r \cos\theta \rangle_\mathrm{sph}}{r} \simeq -\frac{2\pi G}{r^4}\int_{-r}^r \dd z \; z^3 \rho(z).
\end{align} We provide a generalized derivation of this result in Appendix \ref{app:general} where we do not assume $\rho$ is plane-symmetric, and also provide an expression for the error introduced by this approximation.

The quantity on the left-hand side of Equation \ref{eq:the_great_relation} can be related to the vertical number count asymmetry introduced by \cite{Widrow2012}, which has proven a useful tool for studying disk disequilibrium \citep{YannyGardner2013,BennettBovy2019}; see Figure \ref{fig:disk_asym} for an illustration of this idea. We define the residual asymmetry of disk stars to be \begin{equation}
    \delta\rho(z) = \rho(z) - \rho(-z),
\end{equation} which is related to the ``asymmetry parameter'' \begin{equation}
    A(z) = \frac{\rho(z) - \rho(-z)}{\rho(z) + \rho(-z)} = \frac{\delta\rho(z)}{\rho(z) + \rho(-z)}
\end{equation} (although $A(z)$ is usually expressed in terms of number counts of stars, it is simple to convert to density by dividing all terms by the enclosed volume). Since $\delta\rho(z)$ is twice the odd part of $\rho(z)$, Equation \ref{eq:the_great_relation} becomes \begin{equation}
    \dv{\langle a_r \cos\theta \rangle}{r} \simeq -\frac{2\pi G}{r^4}\int_{0}^r \dd z \; z^3 \delta\rho(z).
\end{equation} As $A(z)$ and $\delta\rho(z)$ are often measured in studies of vertical disk waves \citep[or the equivalent $\delta n$, e.g.][]{BennettBovy2019}, this provides a convenient way of translating between observations of accelerations and the density of disk stars. 

Note, however, that this quantity has an additional negative sign compared to $\delta\rho(z)$ and $A(z)$, so a negative value of $\dv*{\langle a_r\cos\theta\rangle}{r}$ corresponds to an enhancement of mass above the midplane.

\begin{figure}
    \centering
    \includegraphics[width=\linewidth]{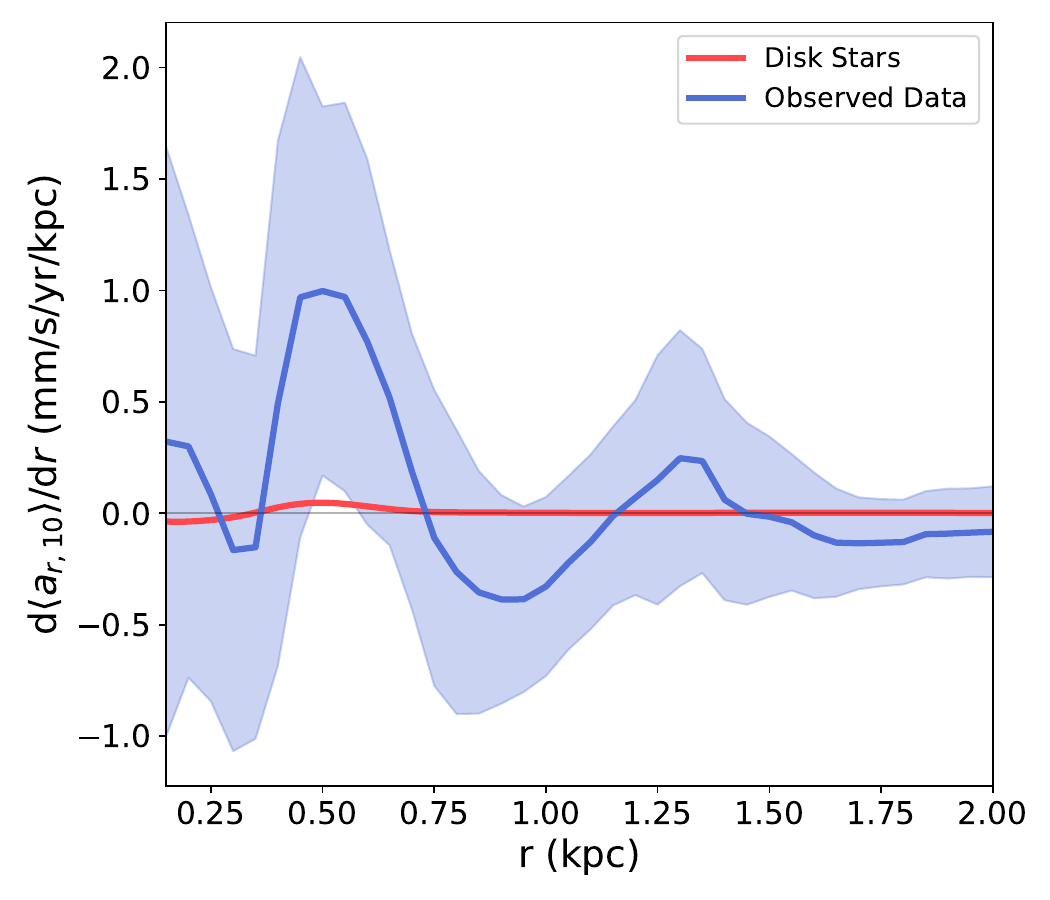}
    \caption{The quantity $\dv*{\langle a_r\cos\theta\rangle}{r}$ as a function of distance from the Sun. This quantity for the observed pulsar data (blue), with a 1$\sigma$ uncertainty region obtained from 100 Monte Carlo draws. Shown in red is this quantity for the observed disk density asymmetry (Figure \ref{fig:disk_asym}). The observed distribution is markedly different from the known disk distribution, which could correspond to the contribution from dark matter. The uncertainties on the observed data are large compared to the purported signal.}
    \label{fig:dar_costheta_dr}
\end{figure}

\subsection{Results}

Figure \ref{fig:dar_costheta_dr} shows the quantity $\dv*{\langle a_r\cos\theta\rangle}{r}$ calculated for the observed pulsar accelerations. We also show the contribution to this quantity from the known disk density distribution; the difference between the observed value and the disk contribution corresponds to additional component, which could arise from asymmetries in the local distribution of dark matter.  

The observed data produces a signal that is much larger than the known asymmetry in the distribution of disk stars. This could correspond to a nonuniform distribution of dark matter on either side of the MW midplane at these distances. The value of $\dv*{\langle a_r\cos\theta\rangle}{r}$ from the pulsar data would imply a density asymmetry of roughly 0.015 $M_\odot/\rm{pc}^3$ near the Sun, which is comparable to the entire purported midplane density of dark matter. It should be reiterated, however, that the uncertainties of the data are large compared to the observed signal, and our results are also consistent with non-detection, or a much smaller signal. More precise data is required before we can make a robust claim about the distribution of dark matter near the Sun. 


One could imagine that such a feature could potentially be caused by gas rather than dark matter, such as a giant molecular cloud located above the Sun \citep[e.g.]{Miville-Deschenes2017}. However, we show in Appendix \ref{sec:app_gas} that the contributions to the density asymmetry from molecular and atomic gas are negligible for the purposes of this work, and as a result cannot produce such a discrepancy between the observed asymmetry and the disk stars. 

\subsection{Signatures of Dark Substructures}

\begin{figure}
    \centering
    \includegraphics[width=\linewidth]{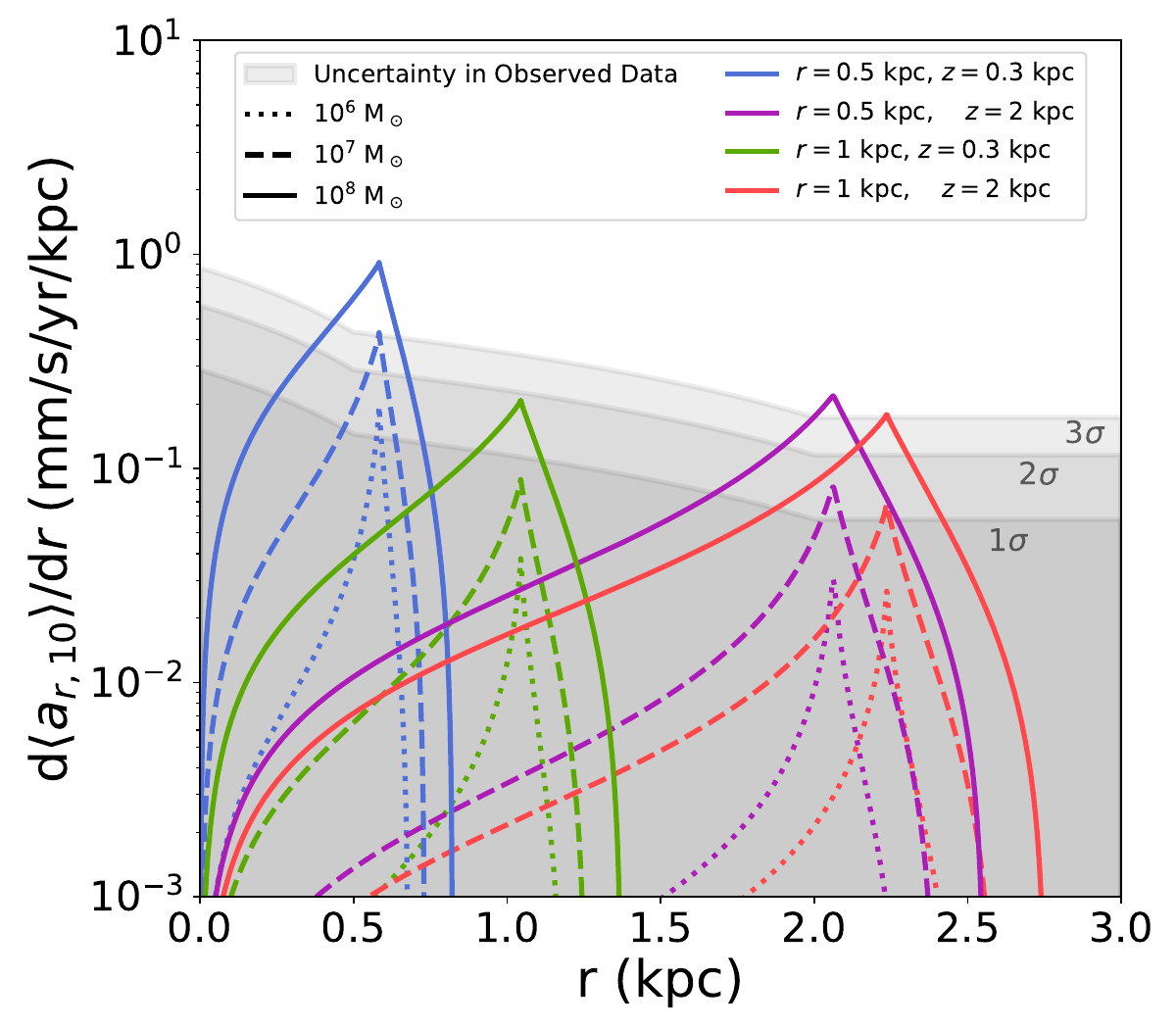}
    \caption{Sensitivity of $\dv*{\langle a_{r,10}\rangle}{r}$ to dark matter subhalos. The theoretical contribution to $\dv*{\langle a_{r,10}\rangle}{r}$ from three subhalo masses are shown as different pattern lines, and colors correspond to different locations relative to the Sun and the Galactic midplane. Uncertainty regions from the observed data are shown in gray; subhalos can currently be detected above these regions. We are currently only able to reliably detect the most massive subhalos ($M\geq10^8$ M$_\odot$), although it may be possible to observe less massive subhalos with low significance, depending on their location relative to the Sun. In general, the signals from subhalos that are closer to the Sun and/or the midplane are stronger than more distant subhalos. }
    \label{fig:dmsh}
\end{figure}

As discussed above, a potential use for this method is the detection of dark substructures near the Sun. This includes dark matter subhalos, relics of cold dark matter models for structure formation \citep{ZavalaFrenk2019}. These structures would be massive enough to gravitationally collect an overdensity of dark matter, but not massive enough to collect the gas required for star formation. This limits their masses from roughly 10$^8$ M$_\odot$ on the upper end, down to $10^4$ or 10$^5$ M$_\odot$ on the lower end.

Recently, \cite{Chakrabarti2025} find constraints at the $\sim 2$--$3\sigma$ level for the mass significance (or Bayes factors of $\sim 20$--40) for a dark matter subhalo near the Sun using pulsar accelerations. In order to provide a comparison to that work, here we perform an analysis of how sensitive our method is to the presence of dark matter subhalos.

We model each subhalo as an NFW profile \citep{NFW}, with a mass of $10^6$, $10^7$, or $10^8$ M$_\odot$. These subhalos have scale radii 0.03, 0.1, and 0.3 kpc respectively, which fall along the subhalo mass-radius relation \citep{ZavalaFrenk2019}. These subhalos are placed some distance $r$ from the Sun and some distance $z$ from the Galactic midplane; then, their contribution to $\dv*{\langle a_{r,10}\rangle}{r}$ is computed from Equation \ref{eq:the_great_relation} (or equivalently, Equation \ref{eq:a_r_10_def}). 

These results are shown in Figure \ref{fig:dmsh}. The contribution of a particular subhalo scales with the mass of the subhalo, and the distance of the subhalo from the Sun sets the location of the peak asymmetry. The peculiar shape of the curve is due to the sign of the line-of-sight acceleration from the NFW profile switching signs at the location of the subhalo, leading to a sharp spike in the asymmetry at the location of the subhalo, which falls off quickly with distance from the subhalo. The lopsided tails on either side of the peak are because $\rm{d}\langle a_{r,10}\rangle/\rm{d}r$ is essentially a volume-weighted average of the acceleration asymmetry; near the Sun, the probed volume is smaller, so the asymmetry falls off less quickly. In summary, more massive subhalos produce larger contributions, and subhalos that are nearby produce larger contributions than those that are farther away. 

With our method, we currently only have the ability to clearly detect massive subhalos with $M\geq10^8$ M$_\odot$ due to the uncertainties in the observed pulsar data. Less massive subhalos could possibly be detected at low significance depending on their position relative to the Sun, although such a detection would be unlikely. It is therefore not particularly surprising that we do not clearly detect the subhalo discussed by \cite{Chakrabarti2025}, which is expected to have a mass of a few times $10^7$ M$_\odot$ and located roughly 1 kpc from the Sun at a height of $z \sim 0.2$ kpc (similar to the dashed purple curve in Figure \ref{fig:dmsh}). Intuitively, it makes sense that our method would have less sensitivity than the \citep{Chakrabarti2025} study; their work uses the accelerations of several pulsars (six binary pulsars and 8 solitary pulsars) located near the location of the subhalo, while our method inherently gets weaker with distance due to the declining number density of pulsars, and averages over the entire sky vs. considering accelerations in a small region of space.  This distinction means that characterizing subhalos from a localized deviation from a smooth potential that is experienced by multiple, nearby pulsars is more likely to yield detections (especially for lower mass subhalos) than searching for an asymmetry in $\dv*{\langle a_{r,10}\rangle}{r}$.  Both in this work and in \cite{Chakrabarti2025}, we do not detect any sub-halos more massive than $10^{8}~M_{\odot}$ within a few kpc of the Sun.

\subsection{Discussion}

Note that the exterior contribution in Equation \ref{eq:a_r_def} has a prefactor of $r^{l-1}$, making it more difficult to remove this contribution for higher-$l$ terms. Removing this external contribution requires taking $l$ derivatives, which will amplify any noise in the observed data. 

An alternative way of thinking about this problem is to consider some measurable $\langle f(\Omega) a_r \rangle$, where $f(\Omega)$ is a weighting function over the sphere. In the recovery of Gauss' Law, we set $f(\Omega) = 1$, whereas in the second example we choose $f(\Omega) = \cos\theta$. Similar choices for the weighting function, such as $\sin\phi$ and $\cos\phi$, correspond to other spherical harmonics (in this case $Y_{1,1}$ and $Y_{1,-1}$), and physically relate to density asymmetries in the Galactocentric $x$ and $y$ directions. 

One might wish to choose, for example, $f(\Omega) = \rm{sgn}(z)$, which would give a ``pure'' measure of the density asymmetry above and below the plane. However, any choice of $f(\Omega)$ can be thought of as a linear combination of the spherical harmonics, and it is impossible to completely remove the external contribution in Equation \ref{eq:a_r_def} from choices such as $\rm{sgn}(z)$, which correspond to an infinite series of spherical harmonics (as one cannot take an infinite number of derivatives of data). 

It is interesting to note that, because $\langle a_r Y_{lm} \rangle$ includes an exterior contribution when $l\geq1$, there is a possibility of using acceleration data as a sort of ``antenna'' to probe distant density distributions. In practice, this is likely difficult; the integrand of the exterior contribution is multiplied by $r^{-(l-1)}$, so the contributions of distant density distributions are dampened rather than amplified.

\section{Conclusion}

We have introduced a novel, non-parametric method to infer the local density distribution of the Milky Way from directly observed acceleration data. By leveraging the divergence theorem, we have related line-of-sight acceleration data for millisecond pulsars to the the mean volume and surface density as a function of distance from the Sun. Since this approach bypasses the assumptions typically made in kinematic analyses -- such as dynamical equilibrium or axisymmetry -- it provides an independent measurement of the Galaxy's density profile.  These results are largely consistent with existing kinematic models. We describe a procedure for determining the mean density of dark matter as a function of distance from the Sun, rather than a single value, which is typically reported by kinematic studies. The mean dark matter appears to decrease with distance, roughly following a power law with a slope of $-0.4\pm1.5$ within 1 kpc of the Sun, and local dark matter density $\rho_{\odot,DM} = 0.006\pm0.012$ $M_\odot$/pc$^3$. However, at this time the relatively large uncertainties on the pulsar acceleration data mean that we are unable to differentiate between a power law and a uniform distribution of dark matter near the Sun.

Additionally, we are able to use higher-order moments of the acceleration field to show the density asymmetry above and below the midplane. The North-South density asymmetry implied by the acceleration data is large compared to the density asymmetry caused by the observed vertical waves in the stellar disk \citep[e.g.][]{BennettBovy2019}; as we cannot explain this feature from the distribution of gas, we suggest that this could potentially be explained by spatial variation in the local density of dark matter. However, current observational uncertainties are large, so at this time we are not able to rule out these feature being caused by noise in the acceleration data. 

The ability to observe asymmetry in density above and below the Galactic midplane further highlights the potential of this method to detect disequilibrium structures that might not be accessible using phase space information of stars alone. This is particularly useful in a case where the potential changes rapidly, such as the flyby of a dark matter subhalo, where the impulse occurs so quickly it is unable to impart significant information onto the motions of stars. This may lead to improved understanding of the MW's local density distribution, particularly our ability to observe dark substructure through the accelerations it produces, and highlights the potential of pulsar acceleration measurements as a tool for advancements in the study of our Galaxy's structure.

While our initial results are promising, the relatively large uncertainty in these measurements currently limits the strength of these claims; with existing data, we are right on the cusp of being able to claim detection of features in the density distribution. Regardless of this limitation, this paper serves as a theoretical basis for how to make use of direct acceleration data in a non-parametric way in order to constrain properties of our Galaxy. As additional pulsar data becomes available -- we in an era where the number of direct acceleration measurements are expected to grow rapidly in the near future -- and as existing pulsar timing solutions are refined, the methods explored in this work are expected to produce more robust results. 

\acknowledgments

We would like to thank Rui Guo for helpful discussion which improved this work.
Sukanya Chakrabarti acknowledges support from NASA EPSCoR CAN AL-80NSSC24M0104, STSCI GO 17505, and the Margaret Burbidge fellowship.
Lawrence Widrow was supported by a Discovery Grant with the Natural Sciences and Engineering Research Council of Canada.


\bibliographystyle{aasjournal}
\bibliography{main.bib}

@ARTICLE{APOGEE,
       author = {{Majewski}, Steven R. and {Schiavon}, Ricardo P. and {Frinchaboy}, Peter M. and {Allende Prieto}, Carlos and {Barkhouser}, Robert and others},
        title = "{The Apache Point Observatory Galactic Evolution Experiment (APOGEE)}",
      journal = {\aj},
         year = 2017,
        month = sep,
       volume = {154},
       number = {3},
          eid = {94},
        pages = {94},
          doi = {10.3847/1538-3881/aa784d},
archivePrefix = {arXiv},
       eprint = {1509.05420},
 primaryClass = {astro-ph.IM},
       adsurl = {https://ui.adsabs.harvard.edu/abs/2017AJ....154...94M}
}

@ARTICLE{Chen2020,
       author = {{Chen}, B.-Q. and {Li}, G.-X. and {Yuan}, H.-B. and {Huang}, Y. and {Tian}, Z.-J. and {Wang}, H.-F. and {Zhang}, H.-W. and {Wang}, C. and {Liu}, X.-W.},
        title = "{A large catalogue of molecular clouds with accurate distances within 4 kpc of the Galactic disc}",
      journal = {\mnras},
         year = 2020,
        month = mar,
       volume = {493},
       number = {1},
        pages = {351-361},
          doi = {10.1093/mnras/staa235},
archivePrefix = {arXiv},
       eprint = {2001.11682},
 primaryClass = {astro-ph.GA},
       adsurl = {https://ui.adsabs.harvard.edu/abs/2020MNRAS.493..351C}
}

@ARTICLE{Craig2025,
       author = {{Craig}, Peter and {Chakrabarti}, Sukanya and {Pettitt}, Alexander R. and {Sanderson}, Robyn and {Rosolowsky}, Erik},
        title = "{A Map of the Outer Gas Disk of the Galaxy with Direct Distances from Young Stars}",
      journal = {\apj},
         year = 2025,
        month = aug,
       volume = {988},
       number = {2},
          eid = {217},
        pages = {217},
          doi = {10.3847/1538-4357/ade232},
archivePrefix = {arXiv},
       eprint = {2506.05575},
 primaryClass = {astro-ph.GA},
       adsurl = {https://ui.adsabs.harvard.edu/abs/2025ApJ...988..217C}
}

@ARTICLE{ZavalaFrenk2019,
       author = {{Zavala}, Jes{\'u}s and {Frenk}, Carlos S.},
        title = "{Dark Matter Haloes and Subhaloes}",
      journal = {Galaxies},
         year = 2019,
        month = sep,
       volume = {7},
       number = {4},
          eid = {81},
        pages = {81},
          doi = {10.3390/galaxies7040081},
archivePrefix = {arXiv},
       eprint = {1907.11775},
 primaryClass = {astro-ph.CO},
       adsurl = {https://ui.adsabs.harvard.edu/abs/2019Galax...7...81Z}
}

@ARTICLE{Chakrabarti2025,
       author = {{Chakrabarti}, Sukanya and {Chang}, Philip and {Profumo}, Stefano and {Craig}, Peter},
        title = "{Constraints on a dark matter sub-halo near the Sun from pulsar timing}",
      journal = {arXiv e-prints},
         year = 2025,
        month = jul,
          eid = {arXiv:2507.16932},
        pages = {arXiv:2507.16932},
          doi = {10.48550/arXiv.2507.16932},
archivePrefix = {arXiv},
       eprint = {2507.16932},
 primaryClass = {astro-ph.GA},
       adsurl = {https://ui.adsabs.harvard.edu/abs/2025arXiv250716932C}
}

@ARTICLE{Buckley2023,
       author = {{Buckley}, Matthew R. and {Lim}, Sung Hak and {Putney}, Eric and {Shih}, David},
        title = "{Measuring Galactic dark matter through unsupervised machine learning}",
      journal = {\mnras},
         year = 2023,
        month = jun,
       volume = {521},
       number = {4},
        pages = {5100-5119},
          doi = {10.1093/mnras/stad843},
archivePrefix = {arXiv},
       eprint = {2205.01129},
 primaryClass = {astro-ph.GA},
       adsurl = {https://ui.adsabs.harvard.edu/abs/2023MNRAS.521.5100B}
}

@ARTICLE{LiWidrow2023,
       author = {{Li}, Haochuan and {Widrow}, Lawrence M.},
        title = "{Residuals of an equilibrium model for the galaxy reveal a state of disequilibrium in the Solar Neighbourhood}",
      journal = {\mnras},
         year = 2023,
        month = apr,
       volume = {520},
       number = {3},
        pages = {3329-3344},
          doi = {10.1093/mnras/stad244},
archivePrefix = {arXiv},
       eprint = {2207.03516},
 primaryClass = {astro-ph.GA},
       adsurl = {https://ui.adsabs.harvard.edu/abs/2023MNRAS.520.3329L}
}

@ARTICLE{LiWidrow2021,
       author = {{Li}, Haochuan and {Widrow}, Lawrence M.},
        title = "{The stellar distribution function and local vertical potential from Gaia DR2}",
      journal = {\mnras},
         year = 2021,
        month = may,
       volume = {503},
       number = {2},
        pages = {1586-1598},
          doi = {10.1093/mnras/stab574},
archivePrefix = {arXiv},
       eprint = {2101.07080},
 primaryClass = {astro-ph.GA},
       adsurl = {https://ui.adsabs.harvard.edu/abs/2021MNRAS.503.1586L}
}

@ARTICLE{Williams2013,
       author = {{Williams}, M.~E.~K. and {Steinmetz}, M. and {Binney}, J. and {Siebert}, A. and {Enke}, H. and {Famaey}, B. and {Minchev}, I. and {de Jong}, R.~S. and {Boeche}, C. and {Freeman}, K.~C. and {Bienaym{\'e}}, O. and {Bland-Hawthorn}, J. and {Gibson}, B.~K. and {Gilmore}, G.~F. and {Grebel}, E.~K. and {Helmi}, A. and {Kordopatis}, G. and {Munari}, U. and {Navarro}, J.~F. and {Parker}, Q.~A. and {Reid}, W. and {Seabroke}, G.~M. and {Sharma}, S. and {Siviero}, A. and {Watson}, F.~G. and {Wyse}, R.~F.~G. and {Zwitter}, T.},
        title = "{The wobbly Galaxy: kinematics north and south with RAVE red-clump giants}",
      journal = {\mnras},
         year = 2013,
        month = nov,
       volume = {436},
       number = {1},
        pages = {101-121},
          doi = {10.1093/mnras/stt1522},
archivePrefix = {arXiv},
       eprint = {1302.2468},
 primaryClass = {astro-ph.GA},
       adsurl = {https://ui.adsabs.harvard.edu/abs/2013MNRAS.436..101W}
}

@ARTICLE{Carlin2013,
	author = {{Carlin}, J.~L. and {DeLaunay}, J. and {Newberg}, H.~J. and 
	{Deng}, L. and {Gole}, D. and {Grabowski}, K. and {Jin}, G. and 
	{Liu}, C. and {Liu}, X. and {Luo}, A.-L. and {Yuan}, H. and 
	{Zhang}, H. and {Zhao}, G. and {Zhao}, Y.},
	title = "{Substructure in Bulk Velocities of Milky Way Disk Stars}",
	journal = {\apjl},
	archivePrefix = "arXiv",
	eprint = {1309.6314},
	year = 2013,
	month = nov,
	volume = 777,
	eid = {L5},
	pages = {L5},
	doi = {10.1088/2041-8205/777/1/L5},
	adsurl = {http://adsabs.harvard.edu/abs/2013ApJ...777L...5C}
}

@ARTICLE{Oort1932,
       author = {{Oort}, J.~H.},
        title = "{The force exerted by the stellar system in the direction perpendicular to the galactic plane and some related problems}",
      journal = {\bain},
         year = 1932,
        month = aug,
       volume = {6},
        pages = {249},
       adsurl = {https://ui.adsabs.harvard.edu/abs/1932BAN.....6..249O}
}

@ARTICLE{Miville-Deschenes2017,
       author = {{Miville-Desch{\^e}nes}, Marc-Antoine and {Murray}, Norman and {Lee}, Eve J.},
        title = "{Physical Properties of Molecular Clouds for the Entire Milky Way Disk}",
      journal = {\apj},
         year = 2017,
        month = jan,
       volume = {834},
       number = {1},
          eid = {57},
        pages = {57},
          doi = {10.3847/1538-4357/834/1/57},
archivePrefix = {arXiv},
       eprint = {1610.05918},
 primaryClass = {astro-ph.GA},
       adsurl = {https://ui.adsabs.harvard.edu/abs/2017ApJ...834...57M}
}

@ARTICLE{HagenHelmi2018,
       author = {{Hagen}, Jorrit H.~J. and {Helmi}, Amina},
        title = "{The vertical force in the solar neighbourhood using red clump stars in TGAS and RAVE. Constraints on the local dark matter density}",
      journal = {\aap},
         year = 2018,
        month = jul,
       volume = {615},
          eid = {A99},
        pages = {A99},
          doi = {10.1051/0004-6361/201832903},
archivePrefix = {arXiv},
       eprint = {1802.09291},
 primaryClass = {astro-ph.GA},
       adsurl = {https://ui.adsabs.harvard.edu/abs/2018A&A...615A..99H}
}

@ARTICLE{Lim2025,
       author = {{Lim}, Sung Hak and {Putney}, Eric and {Buckley}, Matthew R. and {Shih}, David},
        title = "{Mapping dark matter in the Milky Way using normalizing flows and Gaia DR3}",
      journal = {\jcap},
         year = 2025,
        month = jan,
       volume = {2025},
       number = {1},
          eid = {021},
        pages = {021},
          doi = {10.1088/1475-7516/2025/01/021},
archivePrefix = {arXiv},
       eprint = {2305.13358},
 primaryClass = {astro-ph.GA},
       adsurl = {https://ui.adsabs.harvard.edu/abs/2025JCAP...01..021L}
}

@ARTICLE{An2021,
       author = {{An}, J. and {Naik}, A.~P. and {Evans}, N.~W. and {Burrage}, C.},
        title = "{Charting galactic accelerations: when and how to extract a unique potential from the distribution function}",
      journal = {\mnras},
         year = 2021,
        month = oct,
       volume = {506},
       number = {4},
        pages = {5721-5730},
          doi = {10.1093/mnras/stab2049},
archivePrefix = {arXiv},
       eprint = {2106.05981},
 primaryClass = {astro-ph.GA},
       adsurl = {https://ui.adsabs.harvard.edu/abs/2021MNRAS.506.5721A}
}

@ARTICLE{Donlon2025,
       author = {{Donlon}, II, Thomas and {Chakrabarti}, Sukanya and {Widrow}, Lawrence M. and {Vanderwaal}, Sophia and {Ransom}, Scott and {Ramirez-Ruiz}, Enrico},
        title = "{Empirical Modeling of Magnetic Braking in Millisecond Pulsars to Measure the Local Dark Matter Density and Effects of Orbiting Satellite Galaxies}",
      journal = {arXiv e-prints},
         year = 2025,
        month = jan,
          eid = {arXiv:2501.03409},
        pages = {arXiv:2501.03409},
          doi = {10.48550/arXiv.2501.03409},
archivePrefix = {arXiv},
       eprint = {2501.03409},
 primaryClass = {astro-ph.HE},
       adsurl = {https://ui.adsabs.harvard.edu/abs/2025arXiv250103409D}
}

@ARTICLE{SivertssonRead2022,
       author = {{Sivertsson}, S. and {Read}, J.~I. and {Silverwood}, H. and {de Salas}, P.~F. and {Malhan}, K. and {Widmark}, A. and {Laporte}, C.~F.~P. and {Garbari}, S. and {Freese}, K.},
        title = "{Estimating the local dark matter density in a non-axisymmetric wobbling disc}",
      journal = {\mnras},
         year = 2022,
        month = apr,
       volume = {511},
       number = {2},
        pages = {1977-1991},
          doi = {10.1093/mnras/stac094},
archivePrefix = {arXiv},
       eprint = {2201.01822},
 primaryClass = {astro-ph.GA},
       adsurl = {https://ui.adsabs.harvard.edu/abs/2022MNRAS.511.1977S}
}

@ARTICLE{Banik2017,
       author = {{Banik}, Nilanjan and {Widrow}, Lawrence M. and {Dodelson}, Scott},
        title = "{Galactoseismology and the local density of dark matter}",
      journal = {\mnras},
         year = 2017,
        month = feb,
       volume = {464},
       number = {4},
        pages = {3775-3783},
          doi = {10.1093/mnras/stw2603},
archivePrefix = {arXiv},
       eprint = {1608.03338},
 primaryClass = {astro-ph.GA},
       adsurl = {https://ui.adsabs.harvard.edu/abs/2017MNRAS.464.3775B}
}

@ARTICLE{BovyTremaine2012,
       author = {{Bovy}, Jo and {Tremaine}, Scott},
        title = "{On the Local Dark Matter Density}",
      journal = {\apj},
         year = 2012,
        month = sep,
       volume = {756},
       number = {1},
          eid = {89},
        pages = {89},
          doi = {10.1088/0004-637X/756/1/89},
archivePrefix = {arXiv},
       eprint = {1205.4033},
 primaryClass = {astro-ph.GA},
       adsurl = {https://ui.adsabs.harvard.edu/abs/2012ApJ...756...89B}
}

@ARTICLE{Zhang2013,
       author = {{Zhang}, Lan and {Rix}, Hans-Walter and {van de Ven}, Glenn and {Bovy}, Jo and {Liu}, Chao and {Zhao}, Gang},
        title = "{The Gravitational Potential near the Sun from SEGUE K-dwarf Kinematics}",
      journal = {\apj},
         year = 2013,
        month = aug,
       volume = {772},
       number = {2},
          eid = {108},
        pages = {108},
          doi = {10.1088/0004-637X/772/2/108},
archivePrefix = {arXiv},
       eprint = {1209.0256},
 primaryClass = {astro-ph.GA},
       adsurl = {https://ui.adsabs.harvard.edu/abs/2013ApJ...772..108Z}
}

@ARTICLE{Garbari2012,
       author = {{Garbari}, Silvia and {Liu}, Chao and {Read}, Justin I. and {Lake}, George},
        title = "{A new determination of the local dark matter density from the kinematics of K dwarfs}",
      journal = {\mnras},
         year = 2012,
        month = sep,
       volume = {425},
       number = {2},
        pages = {1445-1458},
          doi = {10.1111/j.1365-2966.2012.21608.x},
archivePrefix = {arXiv},
       eprint = {1206.0015},
 primaryClass = {astro-ph.GA},
       adsurl = {https://ui.adsabs.harvard.edu/abs/2012MNRAS.425.1445G}
}

@ARTICLE{RixBovy2013,
       author = {{Rix}, Hans-Walter and {Bovy}, Jo},
        title = "{The Milky Way's stellar disk. Mapping and modeling the Galactic disk}",
      journal = {\aapr},
         year = 2013,
        month = may,
       volume = {21},
          eid = {61},
        pages = {61},
          doi = {10.1007/s00159-013-0061-8},
archivePrefix = {arXiv},
       eprint = {1301.3168},
 primaryClass = {astro-ph.GA},
       adsurl = {https://ui.adsabs.harvard.edu/abs/2013A&ARv..21...61R}
}

@ARTICLE{KuijkenGilmore1991,
       author = {{Kuijken}, Konrad and {Gilmore}, Gerard},
        title = "{The Galactic Disk Surface Mass Density and the Galactic Force K Z at Z = 1.1 Kiloparsecs}",
      journal = {\apjl},
         year = 1991,
        month = jan,
       volume = {367},
        pages = {L9},
          doi = {10.1086/185920},
       adsurl = {https://ui.adsabs.harvard.edu/abs/1991ApJ...367L...9K}
}

@ARTICLE{Cheng2024,
       author = {{Cheng}, Xinlun and {Anguiano}, Borja and {Majewski}, Steven R. and {Arras}, Phil},
        title = "{The surface mass density of the Milky Way: does the traditional K$_{Z}$ approach work in the context of new surveys?}",
      journal = {\mnras},
         year = 2024,
        month = jan,
       volume = {527},
       number = {1},
        pages = {959-976},
          doi = {10.1093/mnras/stad3013},
archivePrefix = {arXiv},
       eprint = {2309.17405},
 primaryClass = {astro-ph.GA},
       adsurl = {https://ui.adsabs.harvard.edu/abs/2024MNRAS.527..959C}
}

@INPROCEEDINGS{CampanaDiSalvo2018,
       author = {{Campana}, Sergio and {Di Salvo}, Tiziana},
        title = "{Accreting Pulsars: Mixing-up Accretion Phases in Transitional Systems}",
    booktitle = {Astrophysics and Space Science Library},
         year = 2018,
       editor = {{Rezzolla}, Luciano and {Pizzochero}, Pierre and {Jones}, David Ian and {Rea}, Nanda and {Vida{\~n}a}, Isaac},
       series = {Astrophysics and Space Science Library},
       volume = {457},
        month = jan,
        pages = {149},
          doi = {10.1007/978-3-319-97616-7_4},
archivePrefix = {arXiv},
       eprint = {1804.03422},
 primaryClass = {astro-ph.HE},
       adsurl = {https://ui.adsabs.harvard.edu/abs/2018ASSL..457..149C}
}

@ARTICLE{HainesDonghia2019,
       author = {{Haines}, Tim and {D'Onghia}, Elena and {Famaey}, Benoit and {Laporte}, Chervin and {Hernquist}, Lars},
        title = "{Implications of a Time-varying Galactic Potential for Determinations of the Dynamical Surface Density}",
      journal = {\apjl},
         year = 2019,
        month = jul,
       volume = {879},
       number = {1},
          eid = {L15},
        pages = {L15},
          doi = {10.3847/2041-8213/ab25f3},
archivePrefix = {arXiv},
       eprint = {1903.00607},
 primaryClass = {astro-ph.GA},
       adsurl = {https://ui.adsabs.harvard.edu/abs/2019ApJ...879L..15H}
}

@ARTICLE{Donlon2024b,
       author = {{Donlon}, Thomas, II and {Chakrabarti}, Sukanya and {Lam}, Michael T. and {Huber}, Daniel and {Hey}, Daniel and others},
        title = "{The Anomalous Acceleration of PSR J2043+1711: Long-Period Orbital Companion or Stellar Flyby?}",
      journal = {arXiv e-prints},
         year = 2024,
        month = jul,
          eid = {arXiv:2407.06482},
        pages = {arXiv:2407.06482},
          doi = {10.48550/arXiv.2407.06482},
archivePrefix = {arXiv},
       eprint = {2407.06482},
 primaryClass = {astro-ph.SR},
       adsurl = {https://ui.adsabs.harvard.edu/abs/2024arXiv240706482D}
}

@ARTICLE{Donlon2024,
       author = {{Donlon}, Thomas, II and {Chakrabarti}, Sukanya and {Widrow}, Lawrence M. and {Lam}, Michael T. and {Chang}, Philip and {Quillen}, Alice C.},
        title = "{Galactic Structure From Binary Pulsar Accelerations: Beyond Smooth Models}",
      journal = {arXiv e-prints},
         year = 2024,
        month = jan,
          eid = {arXiv:2401.15808},
        pages = {arXiv:2401.15808},
          doi = {10.48550/arXiv.2401.15808},
archivePrefix = {arXiv},
       eprint = {2401.15808},
 primaryClass = {astro-ph.GA},
       adsurl = {https://ui.adsabs.harvard.edu/abs/2024arXiv240115808D}
}

@ARTICLE{YannyGardner2013,
       author = {{Yanny}, Brian and {Gardner}, Susan},
        title = "{The Stellar Number Density Distribution in the Local Solar Neighborhood is North-South Asymmetric}",
      journal = {\apj},
         year = 2013,
        month = nov,
       volume = {777},
       number = {2},
          eid = {91},
        pages = {91},
          doi = {10.1088/0004-637X/777/2/91},
archivePrefix = {arXiv},
       eprint = {1309.2300},
 primaryClass = {astro-ph.GA},
       adsurl = {https://ui.adsabs.harvard.edu/abs/2013ApJ...777...91Y}
}

@ARTICLE{BennettBovy2019,
       author = {{Bennett}, Morgan and {Bovy}, Jo},
        title = "{Vertical waves in the solar neighbourhood in Gaia DR2}",
      journal = {\mnras},
         year = 2019,
        month = jan,
       volume = {482},
       number = {1},
        pages = {1417-1425},
          doi = {10.1093/mnras/sty2813},
archivePrefix = {arXiv},
       eprint = {1809.03507},
 primaryClass = {astro-ph.GA},
       adsurl = {https://ui.adsabs.harvard.edu/abs/2019MNRAS.482.1417B}
}

@ARTICLE{Widrow2012,
       author = {{Widrow}, Lawrence M. and {Gardner}, Susan and {Yanny}, Brian and {Dodelson}, Scott and {Chen}, Hsin-Yu},
        title = "{Galactoseismology: Discovery of Vertical Waves in the Galactic Disk}",
      journal = {\apjl},
         year = 2012,
        month = may,
       volume = {750},
       number = {2},
          eid = {L41},
        pages = {L41},
          doi = {10.1088/2041-8205/750/2/L41},
archivePrefix = {arXiv},
       eprint = {1203.6861},
 primaryClass = {astro-ph.GA},
       adsurl = {https://ui.adsabs.harvard.edu/abs/2012ApJ...750L..41W}
}

@BOOK{BinneyTremaine2008,
       author = {{Binney}, James and {Tremaine}, Scott},
        title = "{Galactic Dynamics: Second Edition}",
         year = 2008,
       adsurl = {https://ui.adsabs.harvard.edu/abs/2008gady.book.....B}
}

@ARTICLE{NFW,
       author = {{Navarro}, Julio F. and {Frenk}, Carlos S. and {White}, Simon D.~M.},
        title = "{A Universal Density Profile from Hierarchical Clustering}",
      journal = {\apj},
         year = 1997,
        month = dec,
       volume = {490},
       number = {2},
        pages = {493-508},
          doi = {10.1086/304888},
archivePrefix = {arXiv},
       eprint = {astro-ph/9611107},
 primaryClass = {astro-ph},
       adsurl = {https://ui.adsabs.harvard.edu/abs/1997ApJ...490..493N}
}

@article{Gala,
  doi = {10.21105/joss.00388},
  url = {https://doi.org/10.21105%2Fjoss.00388},
  year = 2017,
  month = {oct},
  publisher = {The Open Journal},
  volume = {2},
  number = {18},
  author = {Adrian M. Price-Whelan},
  title = {Gala: A Python package for galactic dynamics},
  journal = {The Journal of Open Source Software}}

@ARTICLE{Chakrabarti2022,
       author = {{Chakrabarti}, Sukanya and {Stevens}, Daniel J. and {Wright}, Jason and {Rafikov}, Roman R. and {Chang}, Philip and {Beatty}, Thomas and {Huber}, Daniel},
        title = "{Eclipse Timing the Milky Way's Gravitational Potential}",
      journal = {\apjl},
         year = 2022,
        month = apr,
       volume = {928},
       number = {2},
          eid = {L17},
        pages = {L17},
          doi = {10.3847/2041-8213/ac5c43},
archivePrefix = {arXiv},
       eprint = {2112.08231},
 primaryClass = {astro-ph.GA},
       adsurl = {https://ui.adsabs.harvard.edu/abs/2022ApJ...928L..17C}
}

@ARTICLE{Moran2023,
       author = {{Moran}, Abigail and {Mingarelli}, Chiara M.~F. and {Van Tilburg}, Ken and {Good}, Deborah},
        title = "{A Pulsar-Based Map of Galactic Acceleration}",
      journal = {arXiv e-prints},
         year = 2023,
        month = jun,
          eid = {arXiv:2306.13137},
        pages = {arXiv:2306.13137},
          doi = {10.48550/arXiv.2306.13137},
archivePrefix = {arXiv},
       eprint = {2306.13137},
 primaryClass = {astro-ph.GA},
       adsurl = {https://ui.adsabs.harvard.edu/abs/2023arXiv230613137M}
}

@ARTICLE{Buch2019,
       author = {{Buch}, Jatan and {Leung}, John Shing Chau and {Fan}, JiJi},
        title = "{Using Gaia DR2 to constrain local dark matter density and thin dark disk}",
      journal = {\jcap},
         year = 2019,
        month = apr,
       volume = {2019},
       number = {4},
          eid = {026},
        pages = {026},
          doi = {10.1088/1475-7516/2019/04/026},
archivePrefix = {arXiv},
       eprint = {1808.05603},
 primaryClass = {astro-ph.GA},
       adsurl = {https://ui.adsabs.harvard.edu/abs/2019JCAP...04..026B}
}

@ARTICLE{Guo2020,
       author = {{Guo}, Rui and {Liu}, Chao and {Mao}, Shude and {Xue}, Xiang-Xiang and {Long}, R.~J. and {Zhang}, Lan},
        title = "{Measuring the local dark matter density with LAMOST DR5 and Gaia DR2}",
      journal = {\mnras},
         year = 2020,
        month = jul,
       volume = {495},
       number = {4},
        pages = {4828-4844},
          doi = {10.1093/mnras/staa1483},
archivePrefix = {arXiv},
       eprint = {2005.12018},
 primaryClass = {astro-ph.GA},
       adsurl = {https://ui.adsabs.harvard.edu/abs/2020MNRAS.495.4828G}
}

@ARTICLE{McKee2015,
       author = {{McKee}, Christopher F. and {Parravano}, Antonio and {Hollenbach}, David J.},
        title = "{Stars, Gas, and Dark Matter in the Solar Neighborhood}",
      journal = {\apj},
         year = 2015,
        month = nov,
       volume = {814},
       number = {1},
          eid = {13},
        pages = {13},
          doi = {10.1088/0004-637X/814/1/13},
archivePrefix = {arXiv},
       eprint = {1509.05334},
 primaryClass = {astro-ph.GA},
       adsurl = {https://ui.adsabs.harvard.edu/abs/2015ApJ...814...13M}
}

@ARTICLE{BlandHawthornGerhard2016,
       author = {{Bland-Hawthorn}, Joss and {Gerhard}, Ortwin},
        title = "{The Galaxy in Context: Structural, Kinematic, and Integrated Properties}",
      journal = {\araa},
         year = 2016,
        month = sep,
       volume = {54},
        pages = {529-596},
          doi = {10.1146/annurev-astro-081915-023441},
archivePrefix = {arXiv},
       eprint = {1602.07702},
 primaryClass = {astro-ph.GA},
       adsurl = {https://ui.adsabs.harvard.edu/abs/2016ARA&A..54..529B}
}

@ARTICLE{Chakrabarti2020,
       author = {{Chakrabarti}, Sukanya and {Wright}, Jason and {Chang}, Philip and {Quillen}, Alice and {Craig}, Peter and {Territo}, Joey and {D'Onghia}, Elena and {Johnston}, Kathryn V. and {De Rosa}, Robert J. and {Huber}, Daniel and {Rhode}, Katherine L. and {Nielsen}, Eric},
        title = "{Toward a Direct Measure of the Galactic Acceleration}",
      journal = {\apjl},
         year = 2020,
        month = oct,
       volume = {902},
       number = {1},
          eid = {L28},
        pages = {L28},
          doi = {10.3847/2041-8213/abb9b5},
archivePrefix = {arXiv},
       eprint = {2007.15097},
 primaryClass = {astro-ph.GA},
       adsurl = {https://ui.adsabs.harvard.edu/abs/2020ApJ...902L..28C}
}

@ARTICLE{WeisbergHuang2016,
       author = {{Weisberg}, J.~M. and {Huang}, Y.},
        title = "{Relativistic Measurements from Timing the Binary Pulsar PSR B1913+16}",
      journal = {\apj},
         year = 2016,
        month = sep,
       volume = {829},
       number = {1},
          eid = {55},
        pages = {55},
          doi = {10.3847/0004-637X/829/1/55},
archivePrefix = {arXiv},
       eprint = {1606.02744},
 primaryClass = {astro-ph.HE},
       adsurl = {https://ui.adsabs.harvard.edu/abs/2016ApJ...829...55W}
}

@ARTICLE{Shklovskii1970,
       author = {{Shklovskii}, I.~S.},
        title = "{Possible Causes of the Secular Increase in Pulsar Periods.}",
      journal = {\sovast},
         year = 1970,
        month = feb,
       volume = {13},
        pages = {562},
       adsurl = {https://ui.adsabs.harvard.edu/abs/1970SvA....13..562S}
}

@ARTICLE{Chakrabarti2021,
       author = {{Chakrabarti}, Sukanya and {Chang}, Philip and {Lam}, Michael T. and {Vigeland}, Sarah J. and {Quillen}, Alice C.},
        title = "{A Measurement of the Galactic Plane Mass Density from Binary Pulsar Accelerations}",
      journal = {\apjl},
         year = 2021,
        month = feb,
       volume = {907},
       number = {2},
          eid = {L26},
        pages = {L26},
          doi = {10.3847/2041-8213/abd635},
archivePrefix = {arXiv},
       eprint = {2010.04018},
 primaryClass = {astro-ph.GA},
       adsurl = {https://ui.adsabs.harvard.edu/abs/2021ApJ...907L..26C}
}

@ARTICLE{Xu2015,
       author = {{Xu}, Yan and {Newberg}, Heidi Jo and {Carlin}, Jeffrey L. and {Liu}, Chao and {Deng}, Licai and {Li}, Jing and {Sch{\"o}nrich}, Ralph and {Yanny}, Brian},
        title = "{Rings and Radial Waves in the Disk of the Milky Way}",
      journal = {\apj},
         year = 2015,
        month = mar,
       volume = {801},
       number = {2},
          eid = {105},
        pages = {105},
          doi = {10.1088/0004-637X/801/2/105},
archivePrefix = {arXiv},
       eprint = {1503.00257},
 primaryClass = {astro-ph.GA},
       adsurl = {https://ui.adsabs.harvard.edu/abs/2015ApJ...801..105X}
}

@ARTICLE{Antoja2018,
       author = {{Antoja}, T. and {Helmi}, A. and {Romero-G{\'o}mez}, M. and {Katz}, D. and {Babusiaux}, C. and {Drimmel}, R. and {Evans}, D.~W. and {Figueras}, F. and {Poggio}, E. and {Reyl{\'e}}, C. and {Robin}, A.~C. and {Seabroke}, G. and {Soubiran}, C.},
        title = "{A dynamically young and perturbed Milky Way disk}",
      journal = {\nat},
         year = 2018,
        month = sep,
       volume = {561},
       number = {7723},
        pages = {360-362},
          doi = {10.1038/s41586-018-0510-7},
archivePrefix = {arXiv},
       eprint = {1804.10196},
 primaryClass = {astro-ph.GA},
       adsurl = {https://ui.adsabs.harvard.edu/abs/2018Natur.561..360A}
}

@ARTICLE{scipy,
  author  = {Virtanen, Pauli and Gommers, Ralf and Oliphant, Travis E. and
            Haberland, Matt and Reddy, Tyler and Cournapeau, David and
            Burovski, Evgeni and Peterson, Pearu and Weckesser, Warren and
            Bright, Jonathan and {van der Walt}, St{\'e}fan J. and
            Brett, Matthew and Wilson, Joshua and Millman, K. Jarrod and
            Mayorov, Nikolay and Nelson, Andrew R. J. and Jones, Eric and
            Kern, Robert and Larson, Eric and Carey, C J and
            Polat, {\.I}lhan and Feng, Yu and Moore, Eric W. and
            {VanderPlas}, Jake and Laxalde, Denis and Perktold, Josef and
            Cimrman, Robert and Henriksen, Ian and Quintero, E. A. and
            Harris, Charles R. and Archibald, Anne M. and
            Ribeiro, Ant{\^o}nio H. and Pedregosa, Fabian and
            {van Mulbregt}, Paul and {SciPy 1.0 Contributors}},
  title   = {{{SciPy} 1.0: Fundamental Algorithms for Scientific
            Computing in Python}},
  journal = {Nature Methods},
  year    = {2020},
  volume  = {17},
  pages   = {261--272},
  adsurl  = {https://rdcu.be/b08Wh},
  doi     = {10.1038/s41592-019-0686-2},
}

@ARTICLE{GaiaDR3,
       author = {{Gaia Collaboration} and {Vallenari}, A. and {Brown}, A.~G.~A. and {Prusti}, T. and {de Bruijne}, J.~H.~J. and others},
        title = "{Gaia Data Release 3: Summary of the content and survey properties}",
      journal = {arXiv e-prints},
         year = 2022,
        month = jul,
          eid = {arXiv:2208.00211},
        pages = {arXiv:2208.00211},
          doi = {10.48550/arXiv.2208.00211},
archivePrefix = {arXiv},
       eprint = {2208.00211},
 primaryClass = {astro-ph.GA},
       adsurl = {https://ui.adsabs.harvard.edu/abs/2022arXiv220800211G}
}

%
%
%

\appendix

\section{Generalized Derivations} \label{app:general}

In the main text, we provide derivations of the surface density and the vertical density asymmetry that rely on the assumption that the Galactic density profile is plane symmetric, i.e. $\rho = \rho(z)$ only. Here, we provide extensions of those expressions without making this assumption, and also quantify the error associated with each approximation. 

The error in the estimated mean density and surface density due to each approximation is shown for a smooth MW model in Figure \ref{fig:app-approx}. The associated error is small for a potential that is in dynamical equilibrium, and does not significantly impact the results of this work.

\begin{figure}
    \centering
    \includegraphics[width=\linewidth]{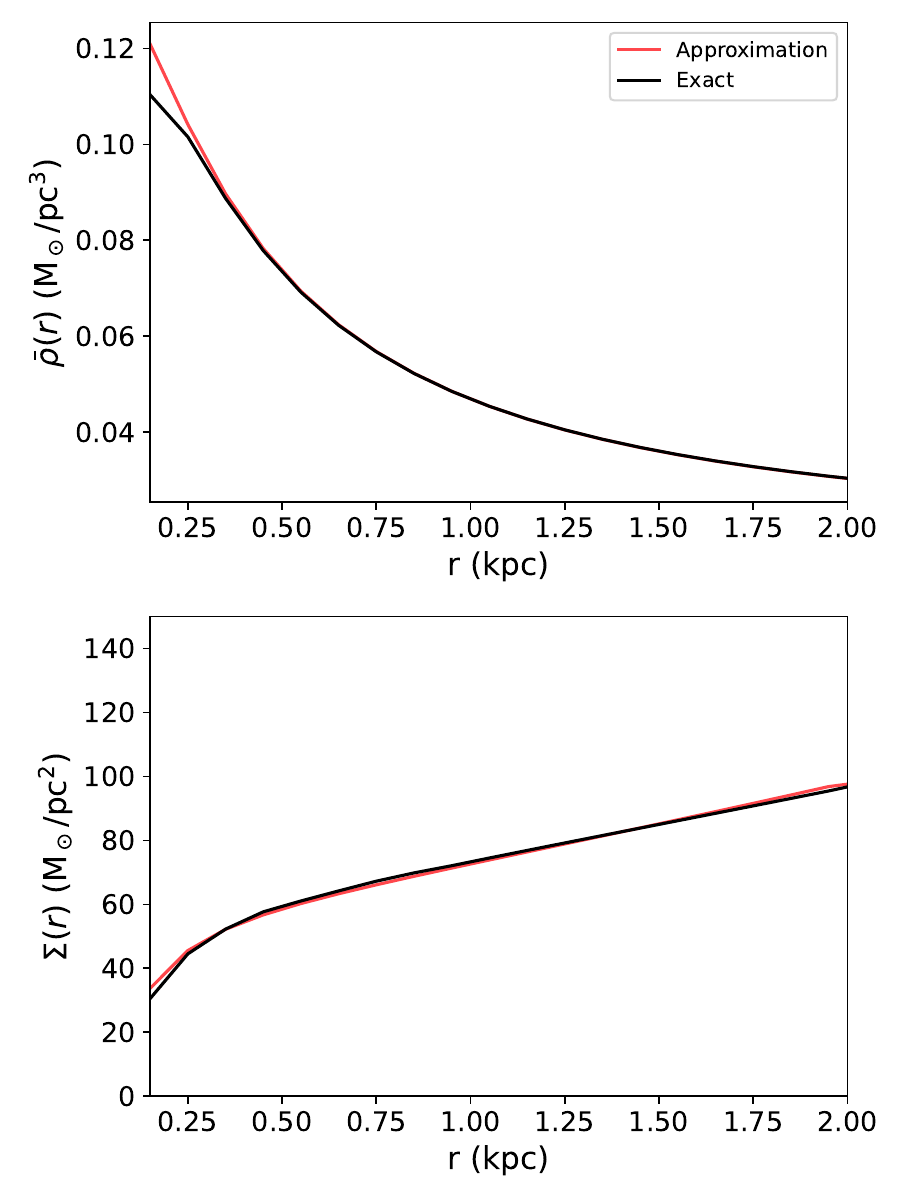}
    \caption{Error in estimates of the mean density (top) and surface density (bottom) due to assumptions of symmetry. In both cases, the differences between the approximation (red) and the true underlying value (black) are small, indicating that the approximations used in this work are acceptable, at least in an equilibrium setting.}
    \label{fig:app-approx}
\end{figure}

\subsection{Surface Density}


We define \begin{equation} \label{eq:sym_adisk}
    \langle \rho\rangle_\mathrm{s,disk}(r,z) = \frac{1}{2A(r,z)}\int_A \dd x \dd y \left( \rho(x,y,z) + \rho(x,y,-z) \right),
\end{equation} as the density averaged over two symmetrically configured disks each at a distance $z$ from the midplane with radius $R=\sqrt{r^2 - z^2}$ and area $A = \pi(r^2 - z^2)$. Here, $x$ and $y$ are the Heliocentric Cartesian coordinates. The ``s'' subscript indicates that we are taking the symmetric part of this quantity with respect to the midplane.

The mean density as a function of distance from the Sun is then \begin{equation}
    \bar{\rho}(r) = \frac{3}{4\pi r^3} \int_0^r \dd z\; \langle \rho\rangle_\mathrm{s,disk}(r,z) A(r,z).
\end{equation} Taking the derivative with respect to $r$, we find \begin{equation} \label{eq:app_drho_dr}
    \dv{\bar{\rho}(r)}{r} = \frac{3}{r^2} \int_0^r  \dd z\;\langle \rho\rangle_\mathrm{s,disk}(r,z) - \frac{3\bar{\rho}(r)}{r}.
\end{equation} 

Surface density is calculated as \begin{equation}
    \Sigma(r,z) = 2\int_0^z \dd z'\; \langle \rho\rangle_\mathrm{s,disk}(r,z').
\end{equation}

The expression for $\Sigma(z)$ in Equation \ref{eq:surf_dens} is recovered in the case where $\langle \rho\rangle_\mathrm{s,disk}(r,z)$ is independent of $r$ (or equivalently, independent of the choice of $R$). In other words, this is to say that $\langle \rho\rangle_\mathrm{s,disk}(r,z) = \bar{\rho}(z),$ or that the density of the disk is not a function of $x$ and $y$. In this case, we can write \begin{align}
    \Sigma(z) &= 2\int_0^z \dd z' \bar{\rho}(z') \\ \nonumber
     &= \int_0^z \dd z' (\rho(0, 0, -z') + \rho(0, 0, z')).
\end{align} The error associated with this approximation is then \begin{align}
    \mathcal{E}(r,z) &= \Sigma(r,z) - \Sigma(z) \\ \nonumber
     &= 2\int_0^z \dd z' \left[ \langle \rho\rangle_\mathrm{s,disk}(r,z') - \langle \rho\rangle_\mathrm{s,axis}(z') \right],
\end{align} where we have introduced the quantity \begin{equation}
    \langle \rho\rangle_\mathrm{s,axis}(z) = \frac{1}{2}\left[\rho(0, 0, -z) + \rho(0, 0, z)\right].
\end{equation} Note that $\mathcal{E}(r,z)$ vanishes when $\rho(x,y,z) = \rho(0,0,z)$ for all $x$ and $y$.

\subsection{Vertical Density Asymmetry}

We begin with the expression for the generalization of the shell theorem from Equation \ref{eq:shell_ext}: \begin{equation}
    \rho_{10}(r) = \sqrt{\frac{3}{4\pi}}\int \dd \Omega \; \rho(r,\Omega) \cos\theta.
\end{equation}

As before, this quantity can be expressed a function of density rather than $\rho_{10}$. We compute the first term on the right-hand side of Equation \ref{eq:tmp3}: \begin{align} \label{eq:app_tmp4}
    \sqrt{\frac{4\pi}{3}}\rho_{10}(r) &= \int_0^{2\pi} \dd \phi \int_{-1}^1 \dd \cos\theta \; \cos\theta \rho(r, \theta,\phi) \\ \nonumber
     &= \frac{2\pi}{r^2} \int_{0}^r \dd z \; z \langle \rho \rangle_\mathrm{a,circ}(r,z),
\end{align} where $\langle \rho \rangle_\mathrm{a,circ}(r,z)$ is the asymmetric part of the mean density over a ring with radius $R=\sqrt{r^2 - z^2}$ and height $z$, and is defined \begin{equation}
    \langle \rho \rangle_\mathrm{a,circ}(r,z) = \frac{1}{4\pi}\int_0^{2\pi} \dd \phi\; \left(\rho(R,\phi,z) - \rho(R,\phi,-z)\right). 
\end{equation} Following a similar procedure, we can write the second term on the right-hand side as \begin{align}
    \sqrt{\frac{4\pi}{3}}\frac{2}{r^4}\int_0^r \dd s \; s^3\rho_{10}(s) = \\ \nonumber
    = \frac{2}{r^4} \int_{-r}^r \dd z \; z \int_{0}^r \dd R\; R \int_0^{2\pi} \dd \phi \; \rho(R,\phi,z) \\ \nonumber
    = \frac{2\pi}{r^4} \int_{0}^r \dd z\; z (r^2 - z^2) \langle\rho\rangle_\mathrm{a,disk}(z),
\end{align} where $\langle\rho\rangle_\mathrm{a,disk}$ is the antisymmetric counterpart to $\langle\rho\rangle_\mathrm{s,disk}$ (Equation \ref{eq:sym_adisk}), and is defined \begin{equation}
    \langle \rho\rangle_\mathrm{a,disk}(z) = \frac{1}{2A(r,z)}\int \dd x \dd y \left( \rho(x,y,z) - \rho(x,y,-z) \right).
\end{equation}

We therefore have \begin{align} \label{eq:app_the_great_relation}
    \dv{\langle a_r \cos\theta \rangle_\mathrm{sph}}{r} = -\frac{2\pi G}{r^4}\int_{-r}^r \dd z \; z^3 \langle \rho\rangle_\mathrm{a,disk}(z) - \mathcal{E}(r),
\end{align} where \begin{equation}
    \mathcal{E}(r) \equiv \frac{2\pi G}{r^2}\int_0^r \dd z\; z \left[ \langle\rho\rangle_\mathrm{a,disk}(z) - \langle\rho\rangle_\mathrm{a,circ}(r,z) \right].
\end{equation} Note that $\mathcal{E}(r)$ vanishes in the case where the density is strictly a function of $z$, so that $\langle\rho\rangle_\mathrm{a,disk}(z) = \langle\rho\rangle_\mathrm{a,circ}(r,z) = \rho(z)$. 

It is intriguing that $\mathcal{E}$ relates the average density across a disk to the average density on the disk's boundary, which is reminiscent of the divergence theorem relating the mean value on the surface of a volume to the mean value interior to that volume, but reduced by one dimension. A similar point can be made about the error in the surface density calculation, which relates the density across a disk to the density averaged across a line; however, the physical implications of this case are less clear, because the relevant line is not the boundary of the disk, but instead the central axis normal to the disk's surface. 

\section{Scaling of the Recovery of the Acceleration Field} \label{sec:app_scaling}

\begin{figure}
    \centering
    \includegraphics[width=\linewidth]{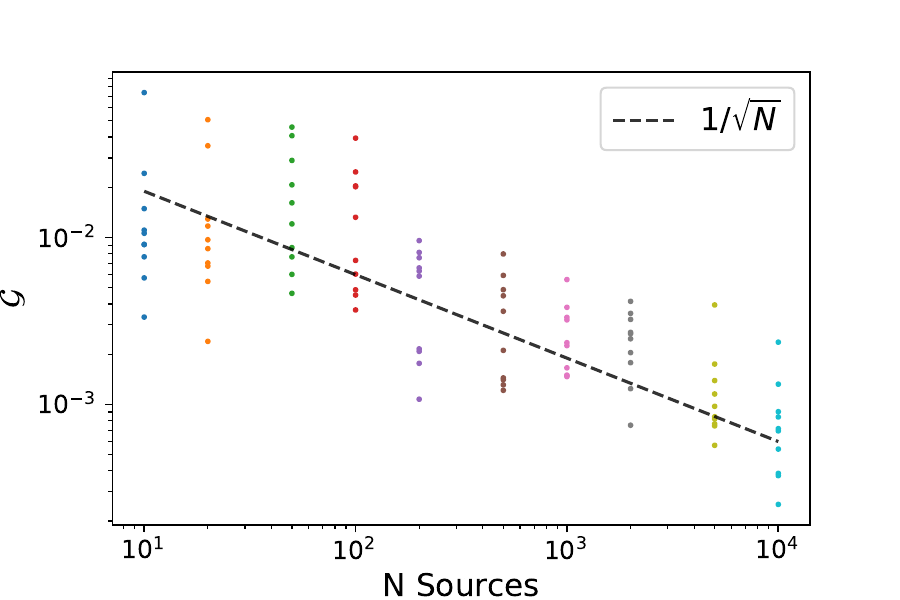}
    \caption{Ability to recover the mean density within 2 kpc of the Sun based on the number of direct acceleration sources used. The quality of the recovered $\bar{\rho}$ distribution scales as $1/\sqrt{N}$, in agreement with Poisson noise. }
    \label{fig:app_scaling}
\end{figure}

Although it is clear from Figure \ref{fig:integral} that increasing the number of courses significantly improves our ability to recover the true underlying line-of-sight acceleration field, it is useful to quantify how quickly this trend scales as a number of accelerometers. To test this, we varied the number of sources between 10 and 10,000 accelerometers, and then computed the quality of the recovery of $\bar{\rho}(r)$ using the goodness-of-fit metric ($\mathcal{G}$) from Section \ref{sec:interp}. This was repeated 10 times for each number of sources we tested to improve the statistics. For each test, we randomly sampled $N$ distances, longitudes, latitudes, and uncertainties from the pulsar dataset, adding artificial noise to each measurement according to the sampled uncertainties. 

Figure \ref{fig:app_scaling} shows the results of this experiment. The quality of the fit trends roughly as $1/\sqrt{N}$, indicating that this procedure is dominated by Poisson noise. However, there is a significant spread in the values of $\mathcal{G}$ across the entire range tested, indicating that the ability of this method to reliably recover the mean density near the Sun depends strongly on the particular conditions of each random draw.

\section{Density Asymmetry of Gas} \label{sec:app_gas}

\begin{figure}
    \centering
    \includegraphics[width=\linewidth]{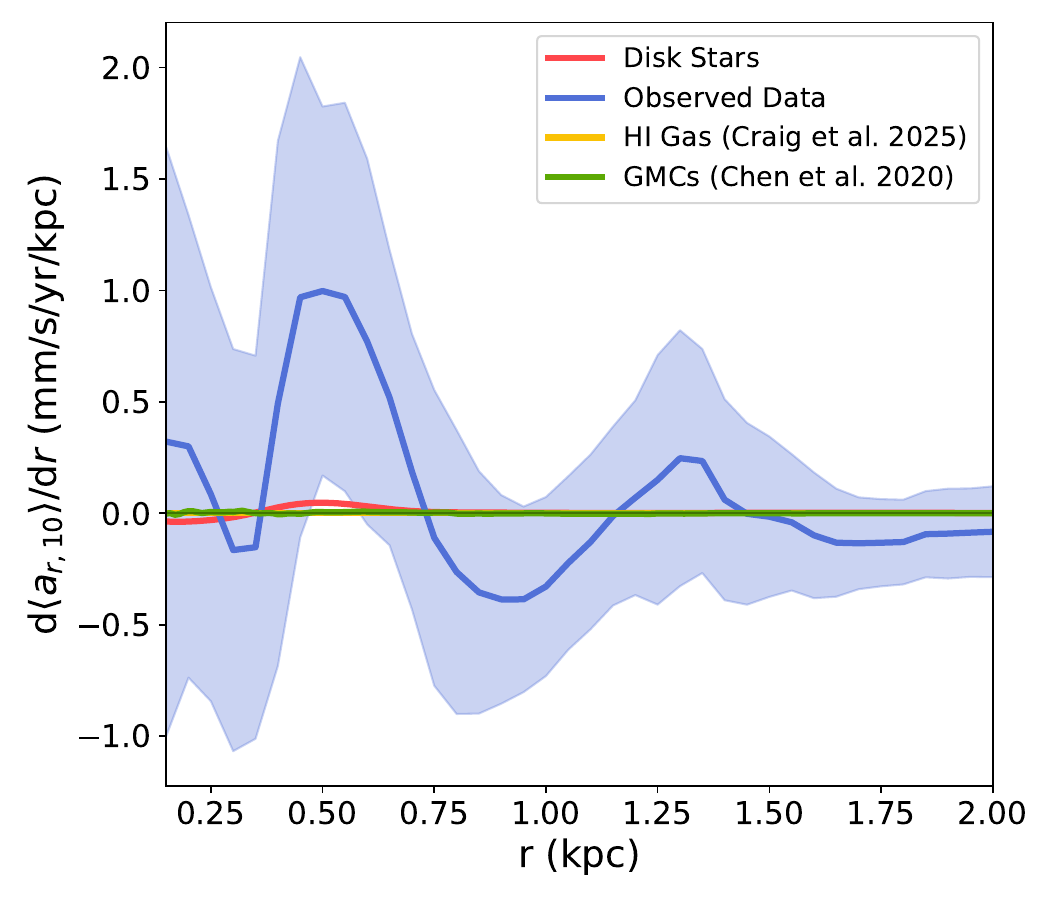}
    \caption{Identical to Figure \ref{fig:dar_costheta_dr}, except we have removed uncertainty regions and also plotted the contributions due to atomic and molecular gas. The effects of gas are small and do not impact the conclusions of this work. }
    \label{fig:app_gas_asymmetry}
\end{figure}

As the distribution of gas throughout the MW is not homogenous, asymmetries in the distribution of gas near the Sun would also be expected to contribute to our measurement of $\dv*{\langle a_{r,10}\rangle}{r}$. If this contribution is substantial compared to the value measured from the observed pulsar accelerations, then the differences between the observed accelerations and those from only stars may simply be due to gas, rather than the local structure of dark matter. 

To test this, we use a three-dimensional map of atomic hydrogen obtained with a novel pattern matching technique \citep{Craig2025}, and a catalog of the positions and locations of giant molecular clouds throughout the Galaxy \citep{Chen2020}. The combination of these two catalogs provides us with a good estimate for the total contribution from atomic and molecular gas in the MW. We then compute $\dv*{\langle a_{r,10}\rangle}{r}$ as a function of $r$ for these catalogs using Equations \ref{eq:a_r_10_def} and \ref{eq:the_great_relation}.

The results are shown in Figure \ref{fig:app_gas_asymmetry}. The overall magnitude of $\dv*{\langle a_{r,10}\rangle}{r}$ for the molecular cloud catalog is $<10^{-2}$ mm/s/yr/kpc, while the contribution from the atomic gas is even smaller at less than $10^{-5}$ mm/s/yr/kpc. We conclude that the distribution of gas does not significantly contribute to our analysis, and that any non-stellar features in the observed accelerations must be due to dark matter or other unknown contributions.  Though the methods are distinct, our conclusions here (that are based on density asymmetries) are consistent with those drawn in \cite{Chakrabarti2025} (that are based on the localized deviation from a smooth potential, as experienced by multiple pulsars) for the contribution of the atomic hydrogen gas to the acceleration field.  

There are two intuitive reasons why the contribution from gas is small. First, gas is collisional, meaning that any vertical density asymmetries are unlikely to be long-lived compared to the vertical distribution of stars. Second, because gas is located very close to the midplane compared to the distribution of stars, the integral on the right-hand side of Equation \ref{eq:the_great_relation} will be very small due to the $z^3$ term in the integrand.

\end{document}